\documentclass[traditabstract,twocolumn]{aa} % for the long lists of affiliations
 \usepackage{graphicx,color}
\usepackage{natbib,twoopt}
\usepackage[breaklinks=true]{hyperref}
\usepackage{pdflscape}

\usepackage{subfigure}

\bibpunct{(}{)}{;}{a}{}{,} % to follow the A&A style
\usepackage{graphicx,latexsym,amssymb}
\usepackage[]{lineno} 

\begin{document}
%\setpagewiselinenumbers
%\linenumbers
\title{Probing  cosmic rays in nearby giant molecular clouds with the Fermi Large Area Telescope}
\author{Rui-zhi Yang\inst{1, 2}
\and Emma de O\~na Wilhelmi\inst{1,5}
\and Felix Aharonian\inst{1, 3, 4}
}
\institute{Max-Planck-Institut f{\"u}r Kernphysik, P.O. Box 103980, 69029 Heidelberg, Germany
\and Key Laboratory of Dark Matter and Space Astronomy, Purple Mountain Observatory, Chinese Academy of Sciences, Nanjing, 210008, China
\and Dublin Institute for Advanced Studies, 31 Fitzwilliam Place, Dublin 2, Ireland
\and Gran Sasso Science Institute, 7 viale Francesco Crispi, 67100 L'Aquila (AQ) , Italy
\and Institut de Cincies de lÕEspai (IEEC-CSIC), E-08193 Bellaterra, Spain
}%
\date{Received  ; accepted }
\abstract{
We report the results of our study of the energy  spectra and absolute fluxes of cosmic rays (CRs)  in the Local Galaxy  based on a five-year  $\gamma$-ray observation with the Fermi Large Area Telescope (LAT)  of  eight nearby giant molecular clouds (GMCs) belonging to the Gould Belt. The $\gamma$-ray  signals obtained  with high statistical significance allow the determination of  $\gamma$-ray spectra  above 300~MeV with adequate  precision for extraction of the energy distributions of CRs in these clouds. Remarkably, both the derived spectral indices and the absolute fluxes of CR protons in the energy interval $10 - 100~ \rm \ GeV$ agree with the recent direct measurements of local CRs by the PAMELA experiment. This is strong evidence of a quite homogeneous distribution of CRs, at least within several hundred parsecs of the Local Galaxy. 
Combined with the well established energy-dependent time of escape of CRs from the Galaxy, $\tau(E) \propto  E^{-\delta}$ with $\delta \approx 0.5-0.6$, the measured spectrum implies a CR spectral index of the (acceleration) source of $\approx$E$^{-2.3}$. At low energies, the spectra of $\gamma$ rays appear to vary from one cloud to another. This implies spatial variations of the energy spectra of CRs below 10~GeV, which at such low energies could be  explained naturally by both the impact of the propagation effects and the contribution of CR locally accelerated inside the clouds. } 

\authorrunning{Yang et al. }
\titlerunning{Giant Molecular Clouds as observed with LAT}

%\titlerunning{}
\maketitle

%%%%%%%%%%%%%%%%%%%%%%%%%%%%%%%%%%%%%%%%%%%%%%

\section{Introduction} 

The current paradigm of cosmic rays (CRs) postulates that the bulk of the CR flux  up to the so-called \emph{knee} around $10^{15}$~eV, is linked to galactic sources, presumably to supernova remnants (cf. Drury 2012 for a recent review). It is also believed that, because of the effective mixture of CRs during their propagation in the interstellar magnetic fields, the CR density locally measured in the Earth's neighbourhood should correctly describe the  average density of CRs throughout the Galactic disk \citep{strong07}, which can be treated as the level of the \emph{sea} of galactic CRs. Small variations of CRs on large (kpc) scales do not, however, exclude significant variations on smaller scales, particularly in the proximity of young CR accelerators. 
Therefore it is not obvious that the locally measured component of CRs can be taken as an undisputed representative of the whole galactic population of relativistic particles. In particular, it is possible that the flux of local CRs might be dominated by the contribution  of a few nearby sources. 
On the other hand, because of the solar modulation effects \citep{pamela2}, the low energy part of the CR flux, typically below 10~GeV, is strongly distorted, thus the direct measurements contain large uncertainties concerning  the level of the \emph{sea} of galactic  CRs at low energies. 
This energy band determines the total luminosity of our Galaxy in CRs and it is crucial for the understanding of several important issues related to the physics of the interstellar medium, e.g. the relative contribution of CRs to the overall pressure in the interstellar medium compared to the magnetic fields and the turbulent and thermal pressure of the gas.  
The interstellar chemistry through the heating and ionisation of the interstellar gas is another important issue related to low energy CRs.  

The density of CRs in different parts of the Galaxy can be  probed uniquely at GeV and hopefully also TeV energies through observations of $\gamma$ rays from massive molecular clouds \citep{FA2001,casanova10,pedaletti}. The realisation of the method requires detailed spectroscopic measurements of a large number of individual giant molecular clouds (GMCs) with known distances, $d$, and masses, $M$. The precision of the method is limited by uncertainties in the ratio of the parameter M/d$^2$ and by the accuracy in the derivation of the spectral shape of CRs, which depends on correctly identifying of the dominant radiation mechanisms and on the accuracy of $\gamma$-ray measurements. The possible presence of potential local accelerators \citep{montmerle79} that can contaminate the CRs \emph{sea} should also be taken into consideration when selecting the GMCs. 

Amongst the best candidates for such studies are GMCs linked to the star formation complexes of the  Gould Belt \citep{FA2001}. These massive clouds (M$\geq$10$^5$M$_\odot$) are located close by (d$<$500~pc) and they are usually located offset of the Galactic plane \citep{gould}, where the large diffuse CRs emission would prevent clear discrimination of the $\gamma$-ray emission associated to clouds at low latitudes. Their location and size make them suitable targets for the Fermi LAT detector \citep{fermi}.

\begin{table*}[!t]
\caption{ Properties of the GMCs investigated in this paper. The numbers in first column  are used to identify the clouds (see Table 2). The estimated distance and position are obtained from Dame et al, 1987. The mass values (dust and CO) listed in the second column are calculated from Planck data and the CfA maps respectively (see text for a description of the mass derivation).} 
\label{tab:1} 
\centering
\begin{tabular}{cccccccc}
\hline\hline
\# & Region & Mass (Dust / CO) [10$^5$ M$_{\odot}$] & Distance [pc] & l [$^{\rm o}$] & b [$^{\rm o}$] & M/d$^2$ [(10$^5$ M$_{\odot}$/kpc$^2)$] & Angular size [${\rm arcdeg}^2$] \\
\hline
1 & $\rho$ Oph &0.12 / 0.08&165&$356^{\circ}$&$ 18^{\circ}$&8.4&68\\
%2 & Hercules  &0.16/0.11 &470&$45^{\circ}$&$ 9^{\circ}$&0.72&18\\
2& Orion B  &0.78 / 0.65 &500 &$205^{\circ}$&-$14^{\circ}$&3.9&22\\
3& Orion A  &1.2 / 0.80 &500&$213^{\circ}$&-$18^{\circ}$&5.2&28\\
4& Mon R2   &1.1 / 0.80  &830&$214^{\circ}$&-$12^{\circ}$&1.7&19\\
5&Taurus  &0.30 / 0.23 &140 &$170^{\circ}$&-$16^{\circ}$&15.0&101\\
6&R CrA &0.01 / 0.01 &150&$0.5^{\circ}$&-$18^{\circ}$&0.8&8\\
7& Chamaeleon & 0.11 / 0.09 &215&$300^{\circ}$&-$16^{\circ}$&2.4&22\\
8& Perseus OB2 &0.41 / 0.3  &350&$158^{\circ}$&-$20^{\circ}$&3.3&27\\
%10& Aquila  &1.1/1.0 &200&$26^{\circ}$&$2^{\circ}$&27.5&92\\
\hline
\end{tabular}
\end{table*}

Recently \citet{nero12} have reported the results of their analysis and interpretation of Fermi LAT observations of a number of GMCs associated with the Gould Belt. In the reduction of data they used a method called  \emph{aperture photometry}\footnote{http://fermi.gsfc.nasa.gov/ssc/data/analysis/scitools/ aperture\_photometry.html}. They claimed evidence of similarity between energy spectra from different clouds. Therefore to derive a global CR spectrum with good statistics, they combine the data from different observed clouds. In this way they conclude that the mean CR spectrum can be described by a steep power-law with an index $3.0\pm 0.2$, with a low energy break at $9 \pm 3~\rm GeV$. However, due to the complex $\gamma$-ray morphology of these objects, to be expected from inhomogeneous spatial distribution of the molecular gas, and (potentially) also CRs, the chosen method for determining energy spectra from these extended objects might lead to misleading conclusions, especially at low energies.   
The Fermi team has also investigated a few representative GMCs \citep{fermiorion,fermimc}. They used the $\gamma$-ray data to determine the calibration ratio between the CO intensity and the column density \citep{dame87,grenier05}, assuming  that the CR spectrum is uniform in the local H~{\sc i} region within Galactocentric radius of $8  10~\rm  kpc$  and in the clouds.
 
Given the importance of these results regarding the origin of galactic CRs, we conducted an independent study on a cloud-by-cloud basis, using a different approach to deriving energy spectra of $\gamma$ rays. Namely, in our analysis we used the \emph{likelihood} method that has been developed and recommended for spectral studies by the Fermi LAT collaboration\footnote{http://fermi.gsfc.nasa.gov/ssc/data/analysis/scitools/ likelihood\_tutorial.html}. 
To compute the cloud density, which is relevant to estimating the CR content, we performed a comparative analysis using data obtained with the Planck satellite (Ade et al, 2011) and the CfA Telescope (Dame et al, 2001).
 
The Gould Belt consists of a concentration of stars forming a ring tited towards the galactic disk by $\approx$20$^{\rm o}$. Several of these clusters are identified as active regions of star formation (e.g. Marraco \& Rydgren 1981, Hillenbrand 1997, Shimajiri et al. 2011). Their location offset the Galactic plane (where the diffuse $\gamma$-ray emission is enhanced), and their proximity to Earth makes them suitable candidates for the study of CR interactions with matter. We would like to point out, however, that active star forming regions can contribute substantially to the acceleration of particles and thus enhance the CR density inside these clouds. If so, this could introduce large uncertainties in the estimate of the flux of galactic CRs and their energy spectrum based on $\gamma$-ray observations of these \emph{active clouds}. On the other hand, the opposite cannot be excluded, i.e. a deficit in the $\gamma$-ray flux compared to the minimum flux expected from GMCs due to interactions of the galactic CRs with the ambient gas. At low energies this could be quite a natural consequence of propagation effects that might prevent the free entrance of CRs into the complex of young stars surrounding the clouds and/or penetration of the particles  deep into the dense cores of the clouds, where the bulk emission of $\gamma$ rays could be produced \citep{FA2001}. For high energy protons, typically above 10~GeV, the impact of propagation effects is dramatically reduced, thus a detection of significant deficit in high energy $\gamma$ rays (produced via $\pi^0$-decay), even from a single cloud, would require a revision of the level of the \emph{sea} of galactic CRs \citep{FA2001}.

\begin{table*}[!]
\caption{Spectral characteristics and statistic test (TS) value of the GMC listed in Table 1 obtained from the LAT data. The individual $\chi^2$/d.o.f. of the spectral representation tested are also quoted with the corresponding probabilities in brackets (see text for more details). } 
\label{tab:2} \centering
\begin{tabular}{cccccccc}
\hline\hline
\# & Region& TS & \shortstack{Flux at 3\,GeV \\ {[}10$^{-9}$GeV$^{-1}$cm$^{-2}$s$^{-1}${]}}  & E$_{b}$ [GeV] &$\chi^2$/d.o.f. (BPL) &$\chi^2$/d.o.f. (KPL) & $\chi^2$/d.o.f. (TPL) \\
\hline
1& $\rho$ Oph &11648 &$ 7.7 \pm 0.8$ &$4.7\pm 2.3$&10.7/9 (0.30)&22.2/11 (0.024) & 13.9/11 (0.24)\\
%2&258 &$ 0.45\pm 0.2$ &$2.4\pm 1.3$& 3.0/9 (0.96)&  5.0/11 (0.93)& 3.2/11 (0.99)  \\
2& Orion B&6107 &$3.0 \pm 0.6$  &$3.6\pm 1.3$& 10.8/9 (0.29)&27.9/11 ($2.3\times 10^{-3}$)& 13.1/11 (0.29)\\
3& Orion A &22021 &$5.9 \pm 0.7$  &$4.3\pm 1.2$& 11.0/10 (0.35)&40.1/12 ($4.9\times 10^{-5}$) & 14.0/12 (0.30) \\
4& Mon R2&1607 &$1.3 \pm 0.2$ &$3.0\pm 0.7$&  10.5/10 (0.39)& 29.4/12 ($3.4\times 10^{-3}$)& 13.4/12 (0.34) \\
5&Taurus&5670 &$9.8 \pm 1.5$ &$4.7\pm 1.5$& 10.5/10 (0.39) &  36.9/12 ($2.3\times 10^{-4}$)& 16.5/12  (0.17) \\
6&R CrA&2315 &$1.2 \pm 0.8$ &$0.9\pm 0.8$&  5.1/9 (0.82)&  7.4/11 (0.76)&  15.0/11 (0.18)\\
7& Chamaeleon&2917&$2.0 \pm 0.5$ &$2.0\pm 0.9$& 9.2/9 (0.42)& 24.0/11 (0.01) & 12.0/11 (0.36) \\
8& Perseus OB2&6410  &$3.8 \pm 0.3$ &$4.9\pm 2.1$&  11.7/10 (0.30)&  20.8/12 (0.05) & 17.3/12 (0.14)\\
%10&1405 &$7.7 \pm 0.7$ &$3.6\pm 1.5$&  11.8/8 (0.16)& 94.0/10  ($9\times 10^{-16}$)& 37.0/10 ($5\times 10^{-5}$)  \\
\hline
\end{tabular}
\end{table*}

The $\gamma$-ray studies at energies above several hundred MeVs have other advantages when compared to lower energies. First, the degradation of the LAT angular resolution at low energies introduces non-negligible uncertainties in the energy spectrum and flux of $\gamma$ rays from the clouds. Secondly, whereas at high energies the contribution to the $\gamma$-ray emission via other channels (bremsstrahlung and inverse Compton scattering of electrons) is not expected to be significant, at energies around 100 MeV primary and secondary electrons can contribute to the $\gamma$-ray flux at a flux level comparable to the contribution of $\pi^0$-decay $\gamma$ rays (see figures and discussion in Appendix \ref{apendix1}), which would smear out the differences in the derived proton spectra.   

To conclude, the detectable fluxes of $\gamma$ rays expected from interactions of CR protons and nuclei with the ambient gas, the lack of other competing $\gamma$-ray production mechanisms in molecular clouds, the effective and (relatively) accurate subtraction of the diffuse $\gamma$-ray background above 1 GeV, and (almost) free propagation of high energy CRs through the GMCs make the latter ideal \emph{detectors} for unbiased studies of the spectral and spatial distributions of CRs in the Local Galaxy at energies above 10~GeV/nucleon. The large exposure time on GMCs accumulated over the several years of LAT continuous monitoring allows a high significant detection of $\gamma$ rays in a broad energy band from a number of GMCs in the Gould Belt, and thus makes the probes of CRs in these objects feasible on a source-by-source basis up to energies of about 1~TeV. 

The paper is structured as follows. In Section~2 we describe the GMCs selected for analysis and their characteristics,  including the density estimation using CO and dust tracers. In Section~3 the results of the analysis of the Fermi LAT observations are presented. In Section~4 we derive the CR spectra and fluxes assuming that $\gamma$ rays are produced in interactions of CR protons and nuclei with the ambient gas, and, finally, in Section~5 we discuss the implications of the results.
 
%%%%%%%%%%%%%%%%%%%%%%%%%%%%%%%%%%%%%%%%%%%%%%%%%%%%%%%%%%%%%%% 
 \section{GMCs:  Mass estimation by means of dust and CO observations}
%%%%%%%%%%%%%%%%%%%%%%%%%%%%%%%%%%%%%%%%%%%%%%%%%%%%%%%%%%%%%%
\label{sec2}

 For our study we selected eight massive clouds identified in the CO galactic survey of \citet{dame01} with the CfA 1.2~m millimetre-wave Telescope (see Table \ref{tab:1}). Most of these GMCs belong or are believed with be associated to the Gould Belt. Only one case, R~Coronae Australis (R~CrA), lying  close to the Galactic centre direction in the opposite direction from the Gould Belt, is certain to be unrelated to it.  Besides the CO data provided by Dame et al. 2001, recent observations with the Planck satellite, which provides opacity maps\footnote{http://irsa.ipac.caltech.edu/data/Planck/release\_1/all-sky-maps} of these clouds \citep{planck}, can be used to derive precise information about the masses and distances to these objects (in Table\ref{tab:1}). All selected GMCs are located at high galactic latitude ($|b|>9^{\circ}$). 

To estimate the total mass contained in the $\gamma$-ray emission region from the dust optical depth maps we used the linear formulae that relate the dust opacity and the column density, i.e.  $N_{\rm H}=\tau_D/(\frac{\tau_D}{N_{\rm H}})^{ref}$ where $N_{\rm H}$ and $\tau_D$ are the gas column density and dust optical depth respectively. The reference dust emissivity measured in low $N_{\rm H}$ regions (which is the case in the GMCs) $(\frac{\tau_D}{N_{\rm H}})^{ref}$ can be found in Table 3 in \citet{planck}. In this work we used the opacity map at $353~\rm GHz$ . The total mass of the cloud can be obtained from the expression,
\begin{eqnarray}
 M_{\rm dust} &=& m_{\rm H}N_{\rm H} A_{\rm cloud} \\ \nonumber
&= &m_{\rm H}\tau_D/(\frac{\tau_D}{N_{\rm H}})^{ref} A_{\rm angular} d^2\\ \nonumber
&=&2.72 \times 10^8 d^2 S_{\rm dust} M_{\odot} ,
\end{eqnarray}
  where $m_H$ is the mass of the hydrogen $A_{\rm cloud}$ and $A_{\rm angular}$ refer to the clouds' physical and angular area,  $d_{\rm kpc}$ is the distance of the cloud in kpc and $S_{\rm dust}$ is the dust optical depth integrated over the angular extent of the cloud in $arcdeg^2$. We choose high density regions in which the opacity is larger than $5 \times 10^{-5}$ as our source templates (for R CrA, in which the density is lower, we use a cut at  $2 \times 10^{-5}$). This corresponds to regions in which the opacity is dominated by the CO contribution as mentioned in Sec.4 in \citet{planck}.

 The CO data measured by \citet{dame01} provides an independent mass estimation.  For that, we use the standard assumption of a linear relationship between the velocity-integrated CO intensity, $W_{\rm CO}$, and the molecular hydrogen column density, N(H$_{2}$) as in \cite{dame01} :
\begin{equation}
N(H_{2})/W_{\rm CO}=(1.8 \pm 0.3) \times 10^{20}~\rm cm^{-2}K^{-1}km^{-1}s^{-1}
\end{equation}
This equation yields 
\begin{equation}
M_{\rm CO} /M_{\odot} = (1200\pm 200) S_{\rm CO} \rm d^2_{kpc}
\end{equation}
where  $S_{\rm CO}$ is the CO intensity integrated over the velocity and angular extent of the cloud in $\rm K ~km~ s^{-1} ~arcdeg^2$. We choose the same regions as the source templates we derived from the dust opacity map described in the last paragraph. The correction for the contribution of heavy elements for a factor of about 1.36 is also considered \citep{allen1973}. The results of the total mass obtained using both the dust map and CO map are summarized in Table 1.  For some clouds our results differed slightly from the estimated masses in the original paper \citep{dame87}. This is because we used only the central and more homogeneous region of the clouds (to avoid systematic errors due to filaments and uneven edges). Recently, in a comprehensive review about CO-H$_{2}$ conversion factor \citep{bolatto13} it was argued that a much larger conversion factor of about $4.8 \times 10^{20}~\rm cm^{-2}K^{-1}km^{-1}s^{-1}$ should be used for the case of Chamaeleon. We therefore adopt this value for Chamaeleon in the calculation below. 

It is also important to note that all clouds are extended objects compared to the LAT angular resolution (up to 0.1$^{\rm o}$ for energy $\approx$10GeV, see Table 1). These large sizes provide an excellent opportunity to explore the clouds' morphology with Fermi LAT, excluding uncertainties related to the localization of the $\gamma$-ray production region. Concerning the impact of uncertainties in $M$ and $d$ in the predicted $\gamma$-ray flux, the expected flux level is proportional to the ratio M/d$^{2}$ which depends only on $\sim S_{\rm CO}$ \citep{casanova10} in the CO observation and  $\tau_D/(\frac{\tau_D}{N_H})^{ref} A_{\rm angular} $ in the dust opacity maps. This minimizes the uncertainties and provides high accuracy on the derivation of CR density in the $\gamma$-ray production region. For CO observation we adopted a \emph{standard} conversion factor with an uncertainty of $17\%$ according to Eq. (2). For very dense clouds with large optical depth in CO line, the relation given by Eq. (3) could underestimate the total mass of hydrogen, in some cases perhaps up to a factor of 2  (see Table 1). On the other hand, linear conversion factor from dust opacity and column density can introduce errors in the mass estimation as discussed above. To account for these we artificially increased our systematic error up to 40\%. To obtain the spatial template from the CO data, we integrated the CO cube in the velocity dimension to get the total CO intensity ($W_{CO}$) map of each cloud. The majority of the clouds are located high above the Galactic plane, thus the CO distribution is reduced to a single peak, implying high accuracy when integrating over the whole velocity range.  \\

 The masses calculated using the two different tracers are listed in Table 1. The ones derived from the dust opacity maps (including not only the mass traced by CO but also H~{\sc i}) are in general larger than the ones obtained through CO observations (up to factor 1.5 in some cases). Besidescannot this contribution, the Planck collaboration reported recently the effect of the so-called \emph{dark gas}, observed when  investigating optical depth of thermal dust emission. This dark gas cannot be traced properly by either CO observations or $21\rm cm$ observation for H~{\sc i}, but still contributes significantly to the total gas column density \citep{planck}. 

In regions with low and high gas column density, however, the correlation between the optical depth for various photometric channels and the total gas column density is linear (e.g. Figure 6 in \citealt{planck}), and only in the range between these two limits does the dust optical depth exceed the linear correlation. This excess is what understood by dark gas. In the clouds we are studying here, the influence of this effect is minimized since we are selecting high density regions, as described in \citealt{planck}. 

 Given that the dust data provide more complete information, avoiding further assumptions on the H~{\sc i} content, we used the templates generated from dust opacity maps as  fiducial templates in the analysis that follows (using templates derived from CO observations as cross checks). 
 
\section{Analysis of the Fermi LAT Data}
\label{section3}

\begin{figure*}
\centering
\includegraphics[width=0.45\linewidth, angle=0]{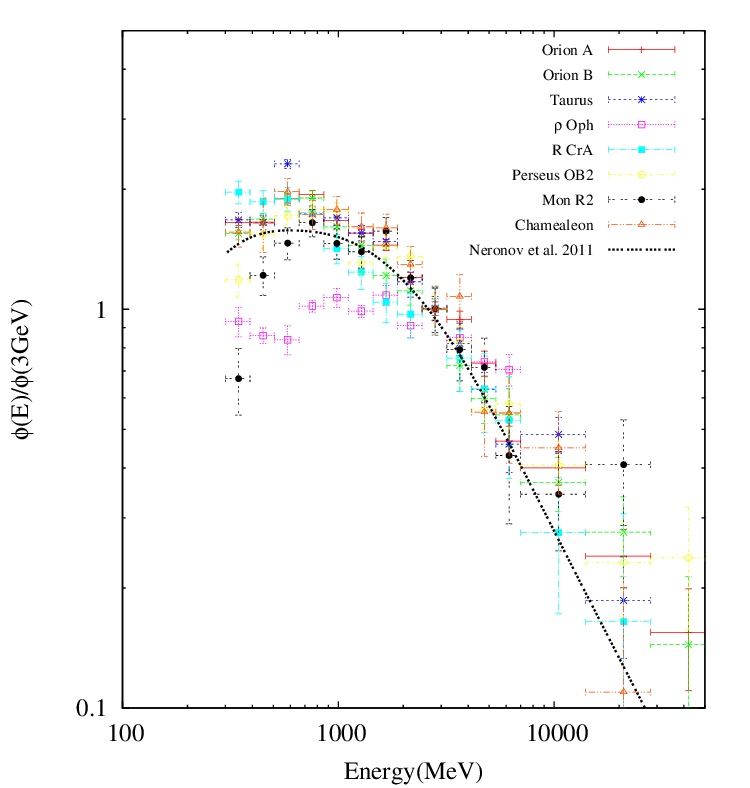}
\includegraphics[width=0.45\linewidth,angle=0]{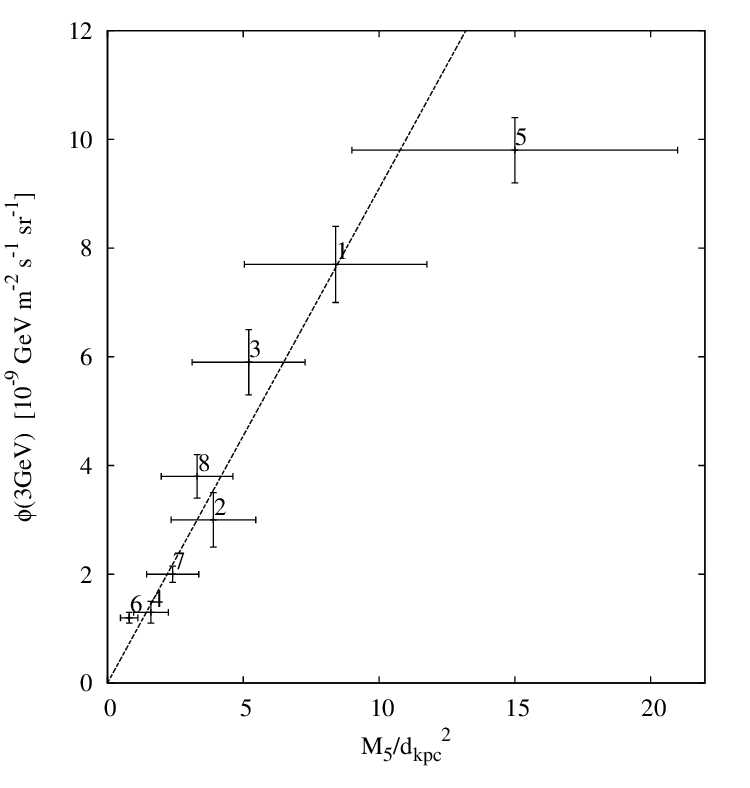}
\caption{{\it Left panel} (a) SED of the $\gamma$-ray emission detected from GMCs in Table~2. For each cloud, the energy spectrum is normalized to its flux at 3\,GeV  to highlight the differences between the spectral shapes derived from different clouds. The spectrum averaged over the $\gamma$-ray data from all clouds reported by \citet{nero12} is also  shown (black curve).  {\it Right panel} (b) Differential $\gamma$-ray flux at 3\,GeV versus $M_5/d_{\rm kpc}^2$ ($M_5$ is the cloud mass in units of $10^5 M_\odot$ and $d_{\rm kpc}$ is distance in units of kpc). The data points are numbered as in Table 2.}
\label{fig:1}
\end{figure*}
%The error bars in $M_5/d_{\rm kpc}^2$ represent $1 \ \sigma$ uncertainty in the conversion factor $N(H$_{2}$)/W_{\rm CO}$ (see text for a more extended discussion)

For the study of the GMCs presented in Table \ref{tab:1} we selected observations with Fermi LAT of the regions around the positions listed in the table integrated for nearly five years (MET 239557417 - MET 380278333). We used the standard LAT analysis software package \emph{v9r31p1}\footnote{http://fermi.gsfc.nasa.gov/ssc}. To avoid systematic errors due to poor angular resolution and uncertainties in the effective detection area at low energies, we selected only events with energy exceeding 300~MeV. The region-of-interest (ROI) was selected to be a $14^ \circ \times 14^ \circ$ square centred on the position of each cloud (as determined by Dame et al. 1987). To reduce the effect of the Earth albedo background, we excluded from the analysis the time intervals when the Earth was in the field-of-view (specifically when the center of the field-of-view was more than $52^ \circ$ from zenith), as well as the time intervals when parts of the ROI had been observed at zenith angles $> 100^ \circ$. The spectral analysis was performed based on the P7v6 version of post-launch instrument response functions (IRFs). Both the front and back converted photons are selected.

Since the $\gamma$ rays produced in GMCs are already included in the galactic diffuse model provided by the Fermi collaboration\footnote{gal\_2yearp7v6\_v0.fit and iso\_p7v6source.txt,
 available at http://fermi.gsfc.nasa.gov/ssc/data/access/lat/BackgroundMo-\\-dels.html}, it cannot be used to evaluate the background. The galactic diffuse emission basically includes the contributions from the inverse Compton (IC) scattering off soft high energy electrons, as well as  $\pi^0$ decay and bremsstrahlung that take place in the H~{\sc i} and H$_{2}$ regions.

As shown in \citet{gabici07}, the contribution of bremsstrahlung emission in passive clouds is expected to be relevant only below 100 MeV, when the electron to proton ratio is $e/p<$0.1. Considering the typical estimated ratio of $e/p\sim\rm 0.01$ as determined in observations of cosmic ray abundances at Earth \citep{hillas05}, bremsstrahlung contribution to the $\gamma$-ray emission can be safely neglected when modelling
passive clouds (see also Appendix \ref{apendix1}). To estimate the background, we calculate only the contributions from IC using GALPROP\footnote{http://galprop.stanford.edu/webrun/} \citep{galprop}, which uses information regarding CR electrons and interstellar radiation field (ISRF).  Isotropic templates related to the cosmic ray contamination and extragalactic $\gamma$-ray background are also included in the analysis. 

To derive the $\gamma$-ray emission from each GMC we used templates generated from dust opacity maps derived by the Planck collaboration \citep{planck}, where we assumed that $\gamma$ rays trace the spatial distribution of the molecular gas.  It should be noted that both the Planck map and CO map have an angular resolution lower than 0.1 degree, which is better than or similar to the counts map of Fermi LAT,  so it can be used as the templates in the Fermi LAT analysis.    Point-like sources from the 2FGL catalogue \citep{2fgl} that appear within the ROI were also included in the analysis. To perform the likelihood analysis, we defined inner circle centred at the ROI centre with a radius of $5^{\circ}$.  The normalization of the point sources inside the inner circle and all the diffuse templates were left free in the analysis.  The parameters of the point sources outside the inner circle were fixed to the catalogue value.   We applied \emph{gtlike} and a binned likelihood analysis to obtain spectral features and test-statistics (TS) values for each clouds (see Table \ref{tab:2}).\\

\begin{figure*}
\centering
\subfigure[][Orion A]{
\includegraphics[scale=0.22]{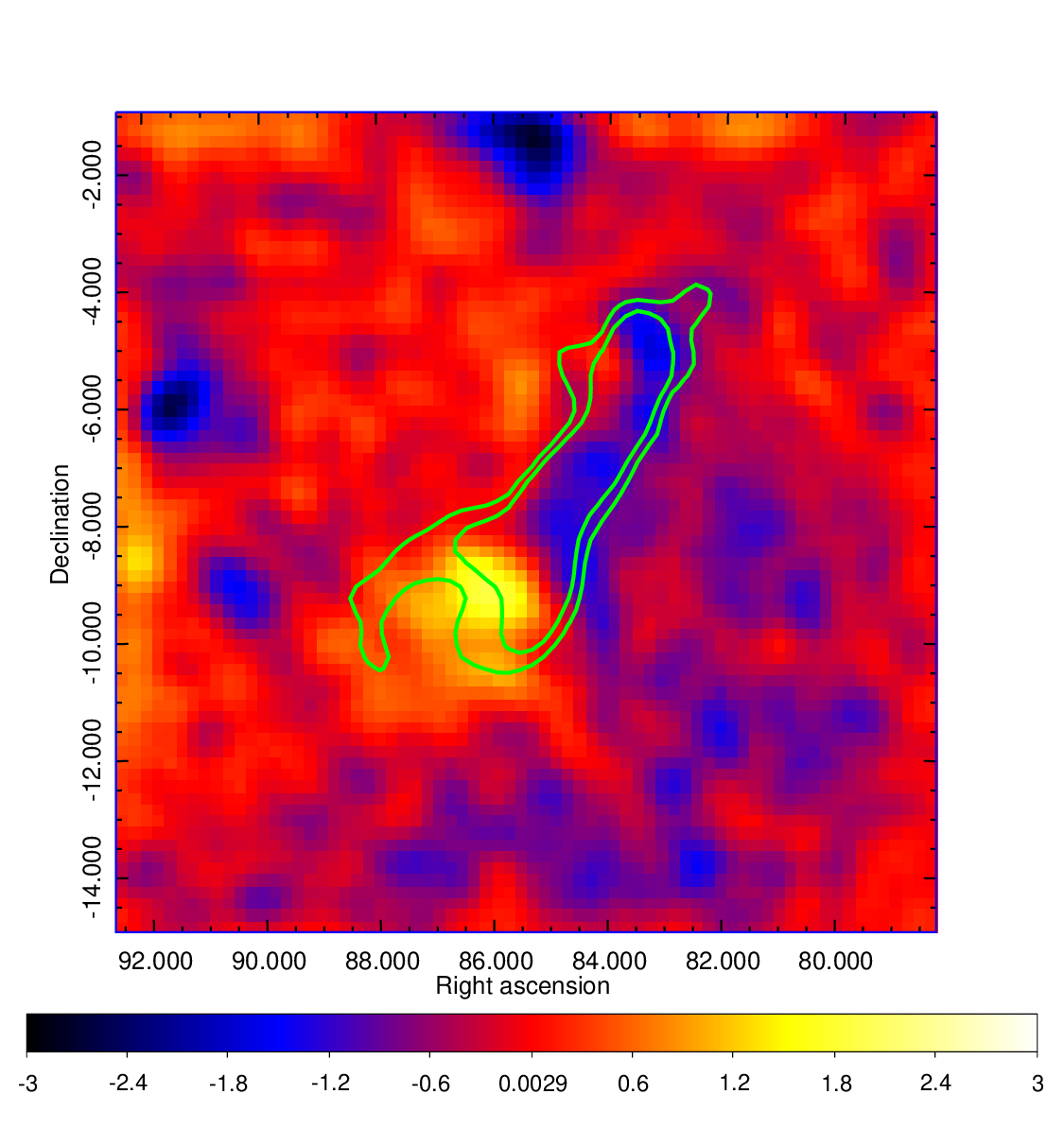}
}
\subfigure[][Orion B]{
\includegraphics[scale=0.22]{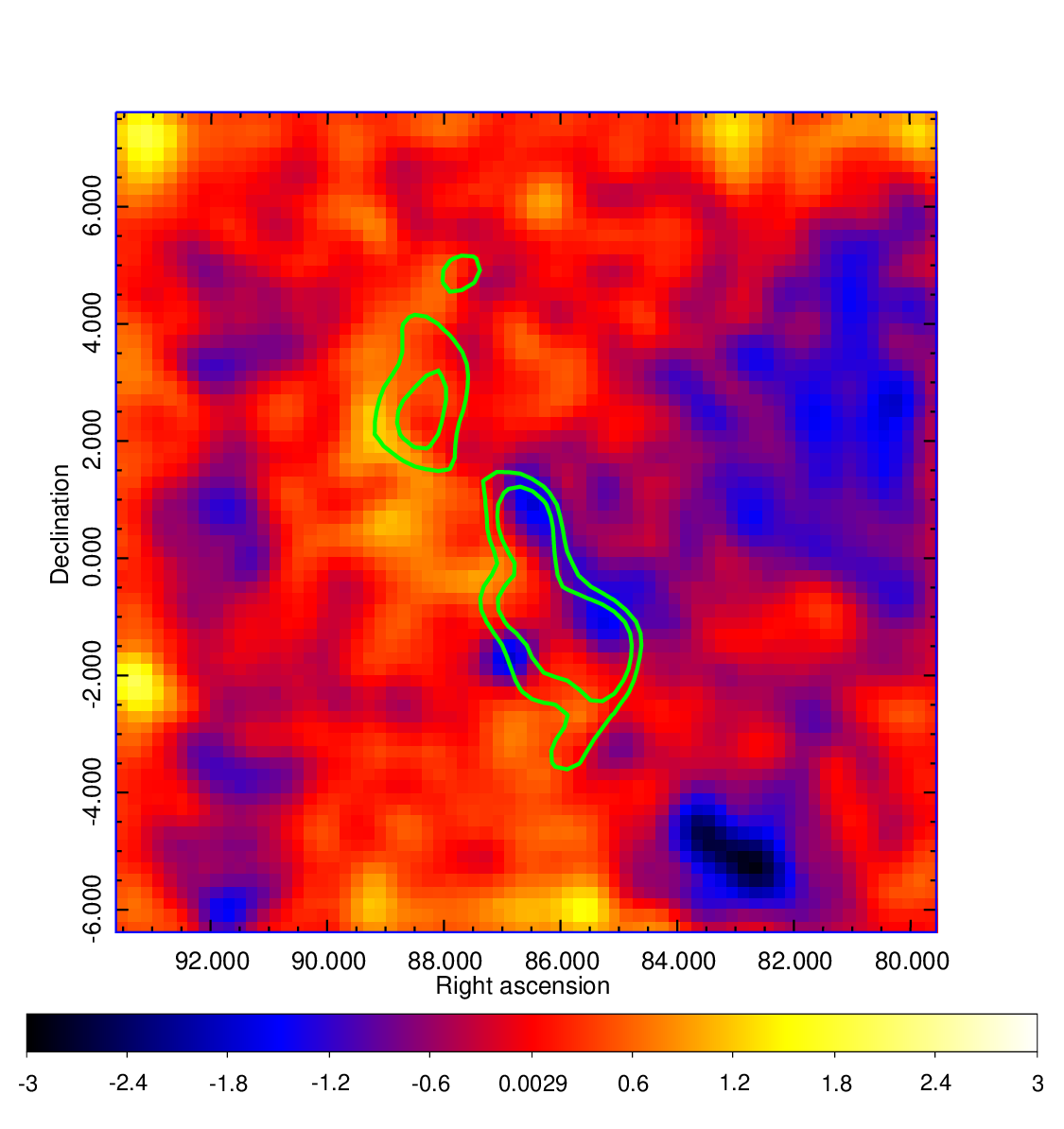}
}
\subfigure[][Mon R2]{
\includegraphics[scale=0.22]{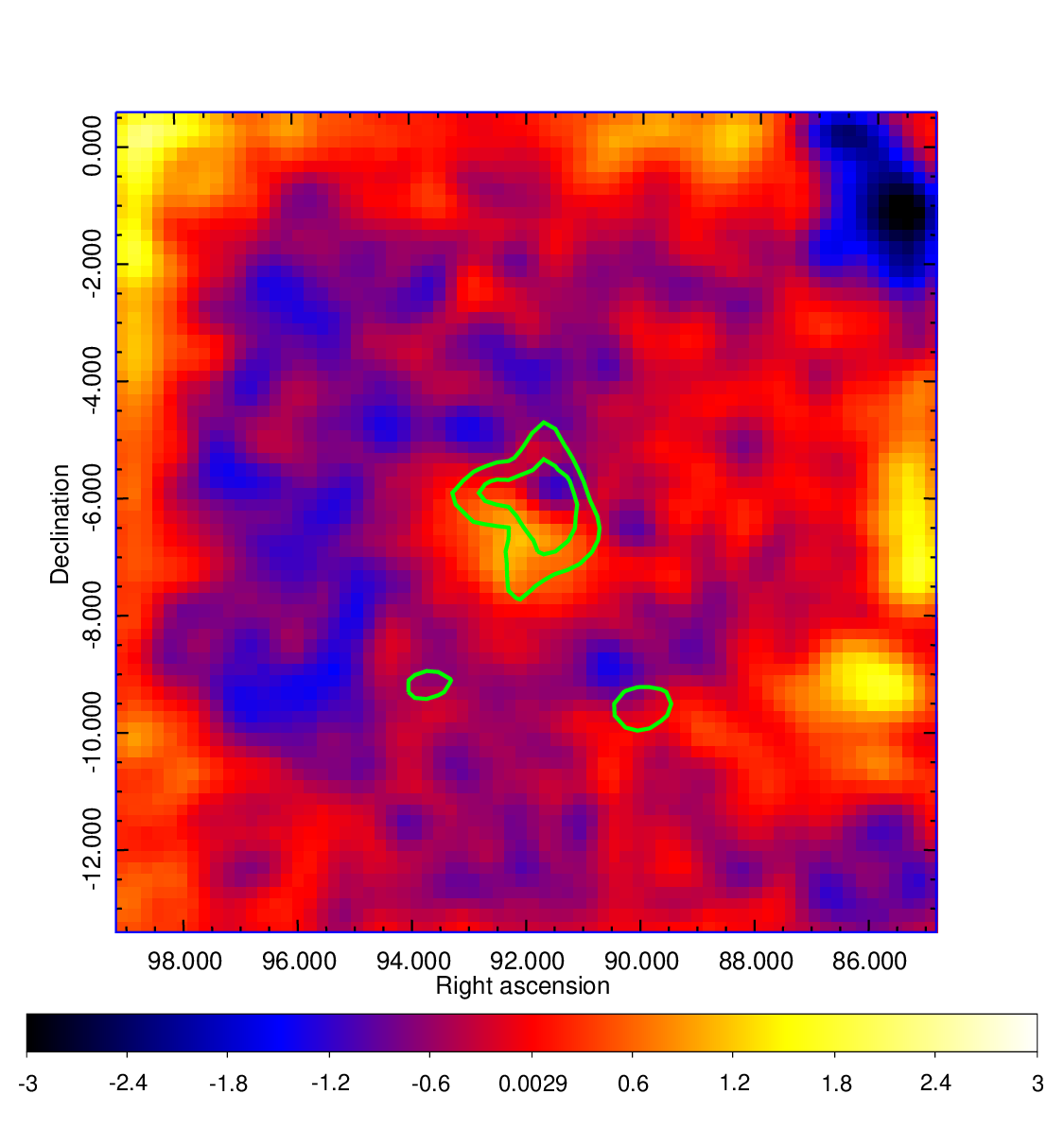}
}
\subfigure[][$\rho$ Op]{
\includegraphics[scale=0.22]{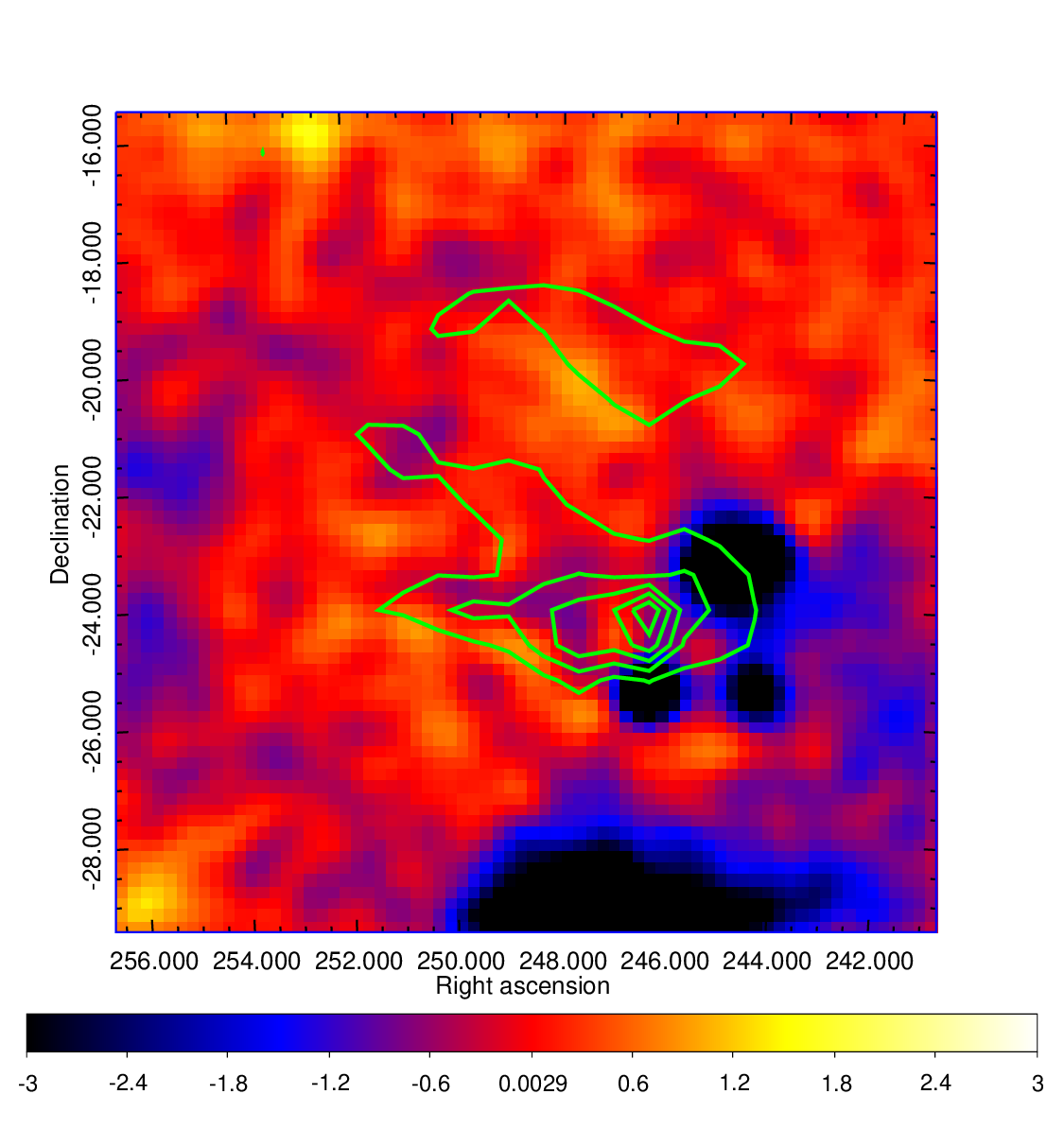}
}
\subfigure[][Perseus OB2]{
\includegraphics[scale=0.22]{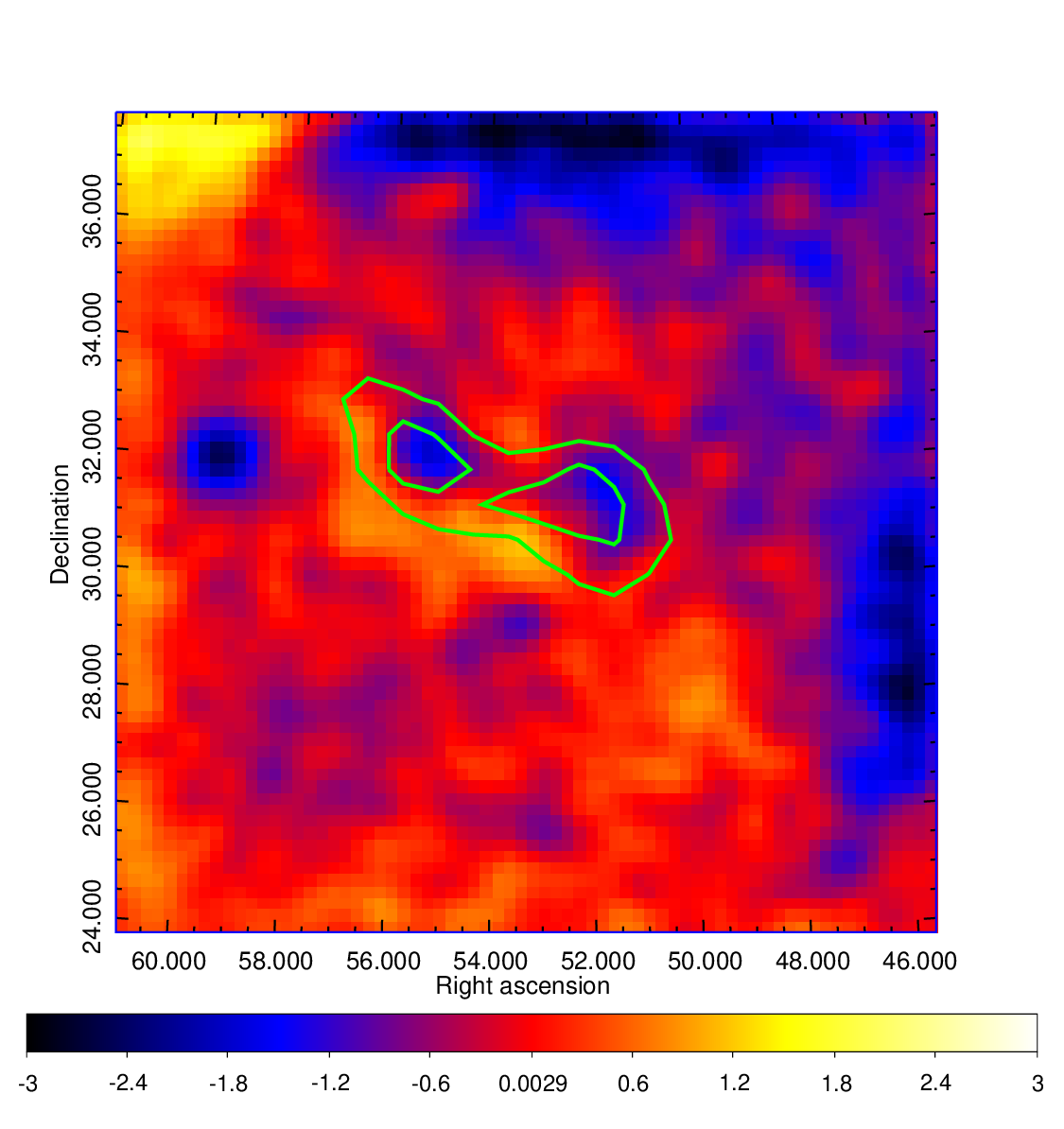}
} 
\subfigure[][Taurus]{
\includegraphics[scale=0.22]{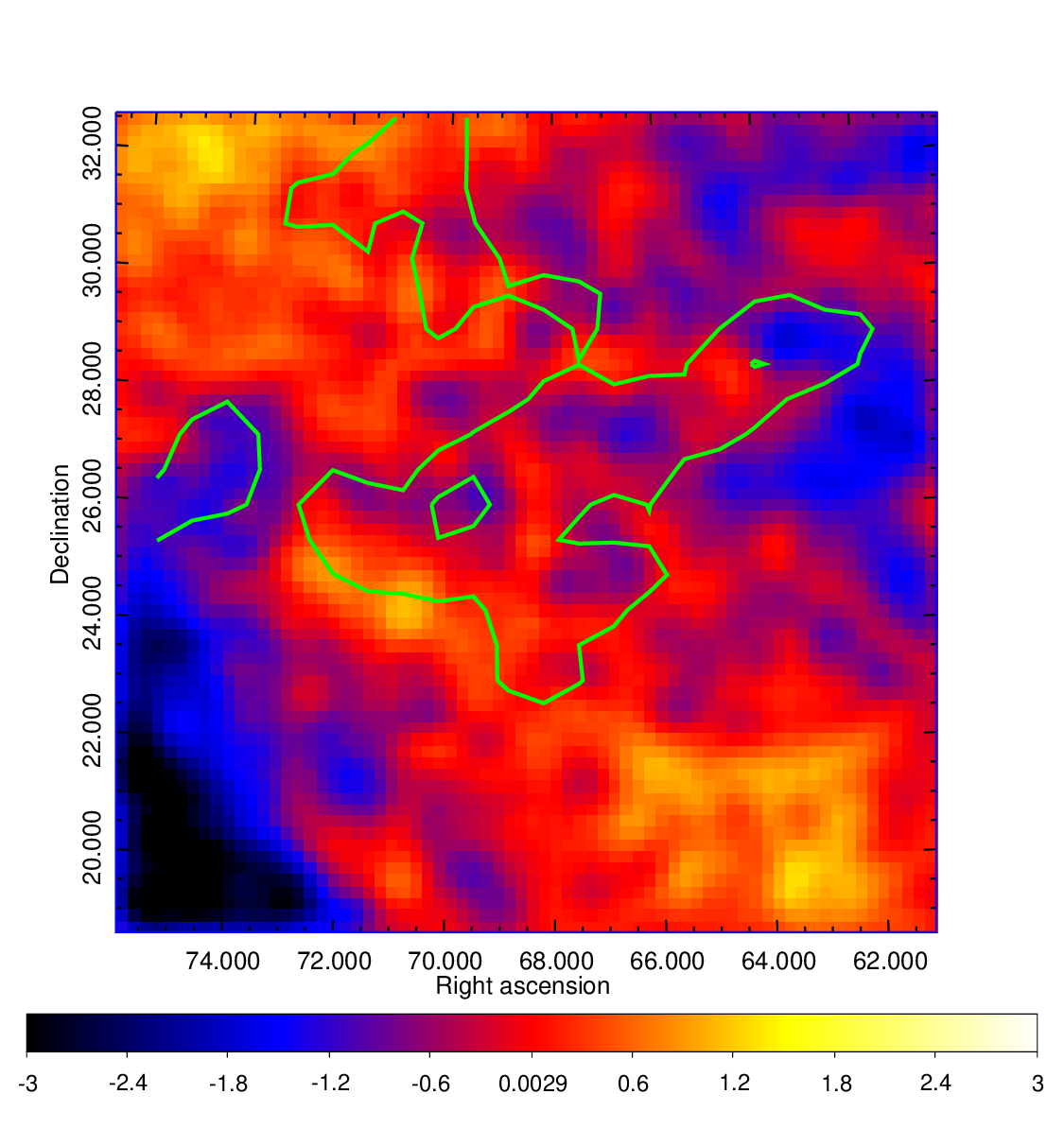}
}
\subfigure[][Chamaeleon]{
\includegraphics[scale=0.22]{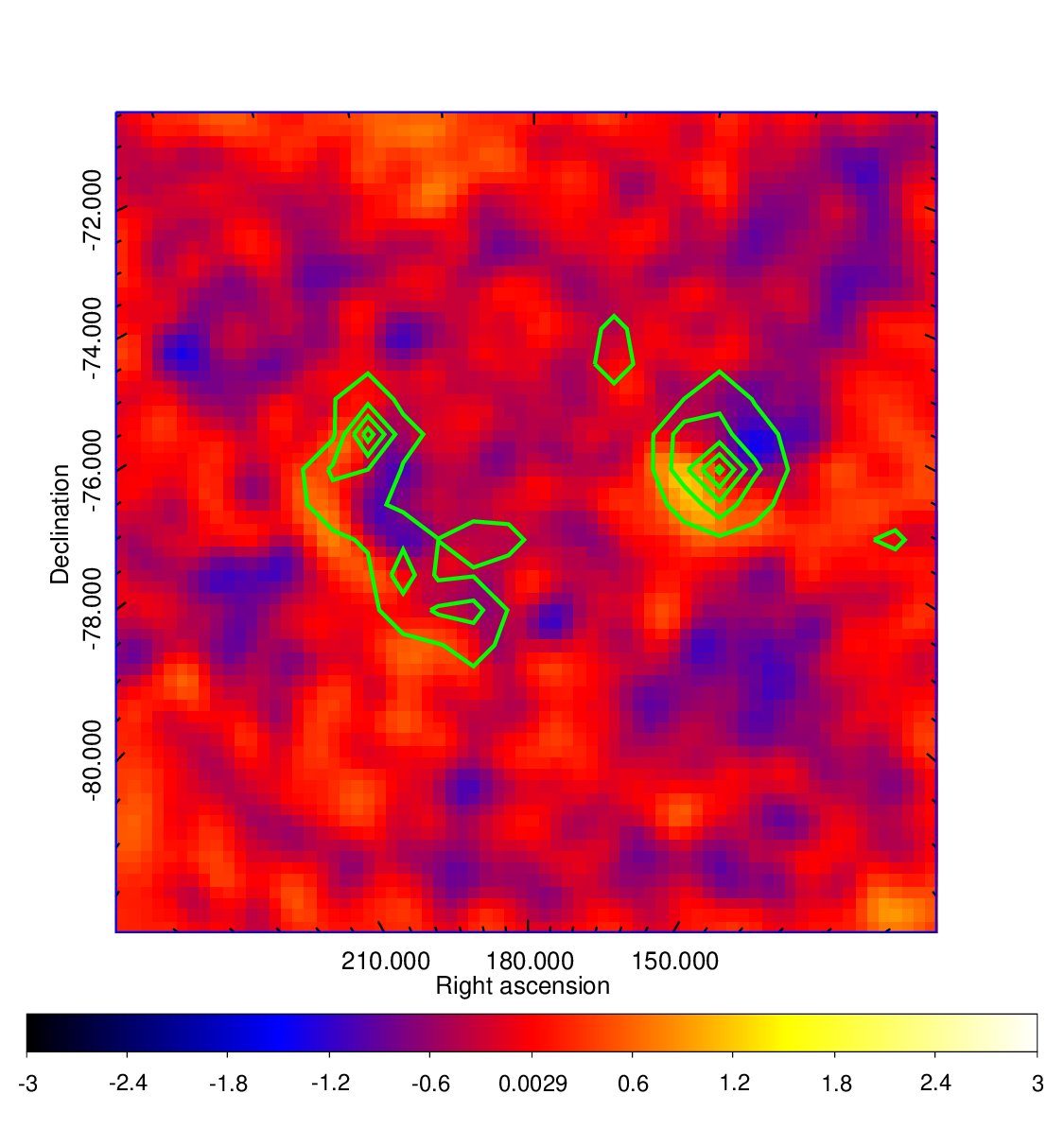}
}
\subfigure[][R CrA]{
\includegraphics[scale=0.22]{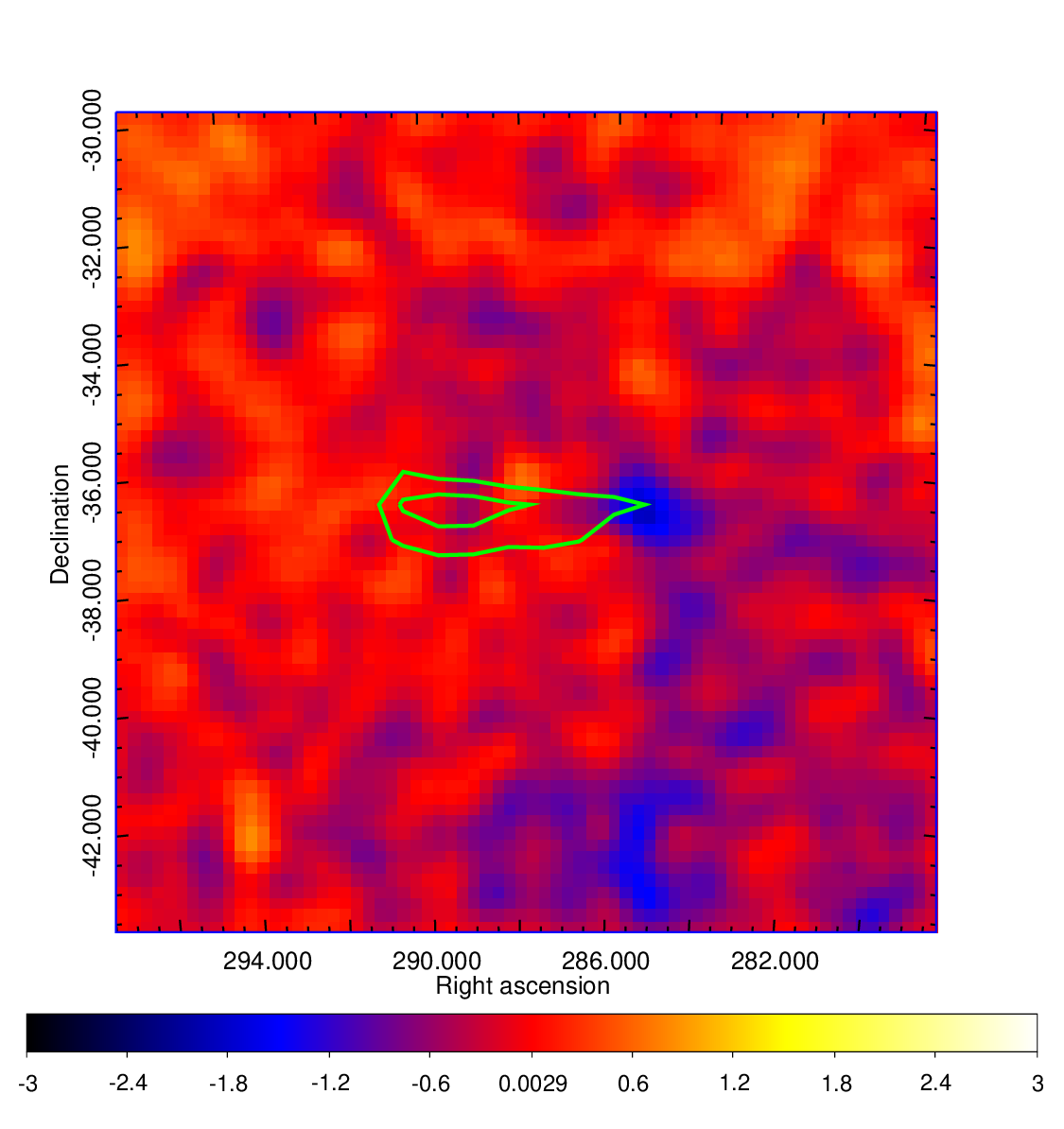}
}
 \caption{Residual S/N maps for all 8 ROIs, scaled to $\pm 3$ standard deviation. The contours represent the regions in which the dust opacity is larger than $5\times10^{-5}$ in the clouds. These correspond to the  source templates described in the text.   } 
\label{fig:res}
\end{figure*}

 The residual signal-to-noise (S/N) maps, after fitting the $\gamma$-ray emission using the templates derived from the clouds, are shown in Fig.\ref{fig:res}.  These maps were derived by dividing the residual counts map with the square root of the raw counts map. All the maps are scaled from $-3$ to $3$, corresponding to $\pm 3$ standard deviation. For seven out of the eight maps, all the residuals are within $\pm 3$ standard deviation. The only exception is $\rho$~Oph, which is overlaid with several strong point sources. Thus the spectral energy distribution (SED) we derived for this cloud is less reliable. 

All eight GMCs included in this study appeared to be strong $\gamma$-ray emitters; they have been detected at a high confidence level with statistical significances TS$>1000$.  To derive their SED we divided the energy range $300~ {\rm MeV} - 60~{\rm GeV}$ into 15 logarithmically spaced bands and applied \emph{gtlike} to each of these bands.  The fitted spectrum of the individual clouds is shown in Figure \ref{fig:1}a, where the SEDs are normalized to the corresponding fluxes at 3\,GeV in order to highlight the differences in the spectral features of the $\gamma$-ray radiation from different clouds. We also show the spectrum reported by Neronov et al (2012) based on the combined $\gamma$-ray data of all clouds.  
This curve can be interpreted as the average energy spectrum. However while at high energies the spectra of individual clouds (see Figure 1) are similar to this one, below $\sim$3 GeV the spectra of some clouds deviate substantially.
  
 Apart from the systematic errors of the LAT's effective area, which were already taken into account in the derived SED, the main source of systematic errors in the $\gamma$-ray spectrum comes from the uncertainty of the chosen diffuse background templates.  For the dust templates, since the dust opacity map does not have the velocity information, the derived mass is integrated over the whole line of sight. As a result the mass of the cloud may be overestimated because H~{\sc i} gas is distributed diffusively in the galaxy. For CO templates the H~{\sc i} and CO map may not be sufficient to model the gas distribution in some regions due to the existence of dark gas.  On the other hand, the spatial templates of IC generated from GALPROP may be model-dependent. To assess the uncertainties introduced by the IC spatial templates, we modified them attempting to fit the results under different hypothesis  (e.g. a model representing the IC contribution as an unrealistic simple disk shape and a second one isotropically filling the field of view). 
Below 1 GeV, this results in a maximum change of $10\%$ to the flux level, while it becomes negligible at high energies. %This is the case for clouds away from the galactic plane whereas in the case of Aquila for instance, changes of the diffuse background might alter the derived $\gamma$-ray flux significantly. Considering that the results presented here should be taken with caution for this particular cloud.  

\begin{figure*}
\centering
\subfigure[][Orion A]{
\includegraphics[scale=0.3]{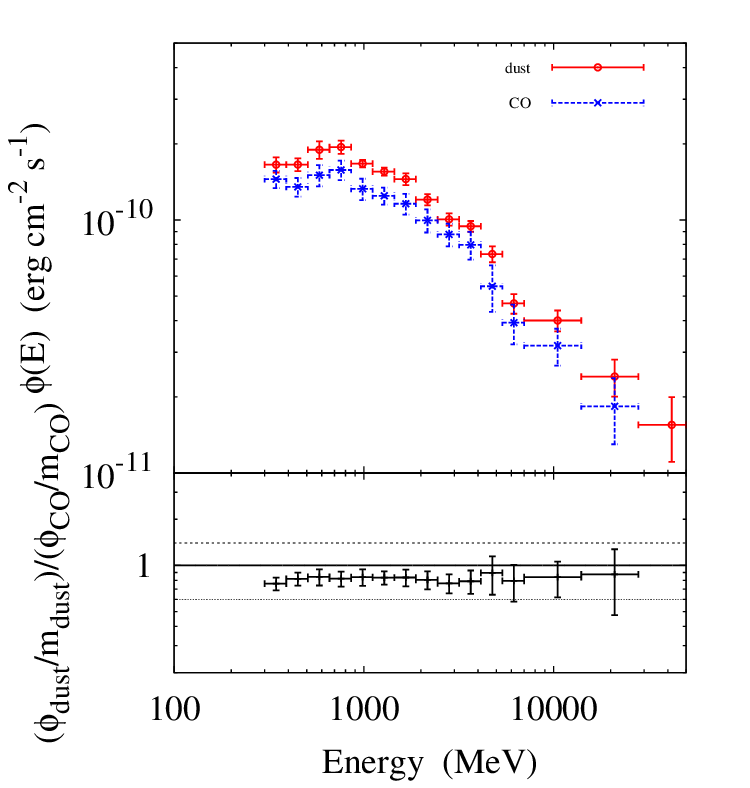}
\label{OrionA:2}
}
\subfigure[][Orion B]{
\includegraphics[scale=0.3]{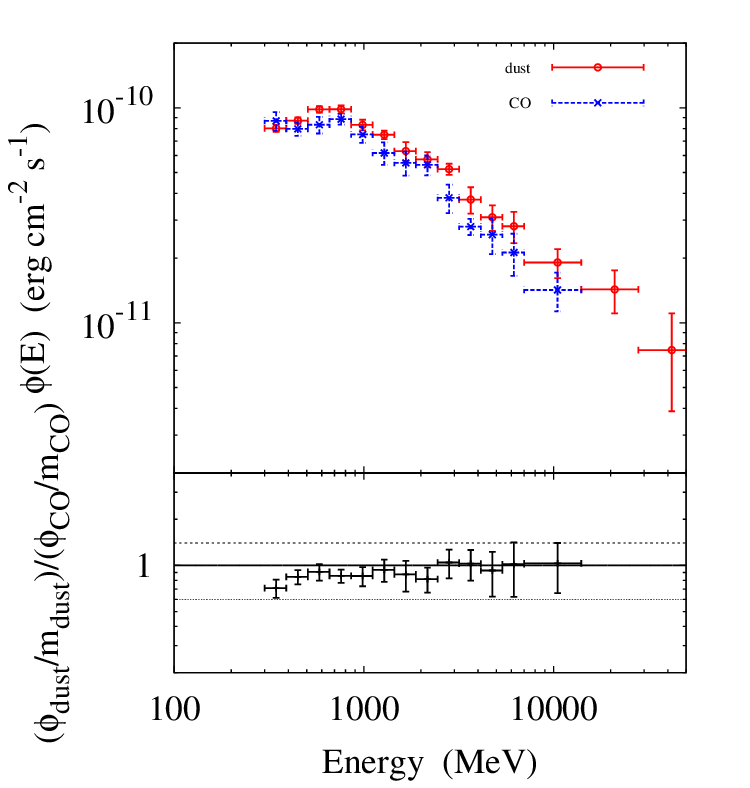}
\label{OrionB:2}
}
\subfigure[][Mon R2]{
\includegraphics[scale=0.3]{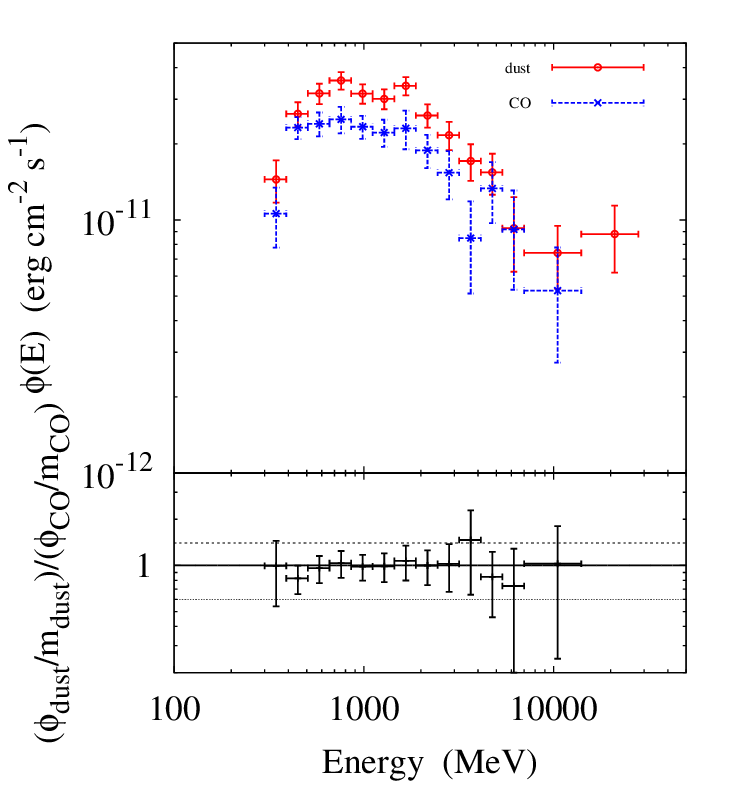}
\label{MonR2:2}
}
\subfigure[][$\rho$ Op]{
\includegraphics[scale=0.3]{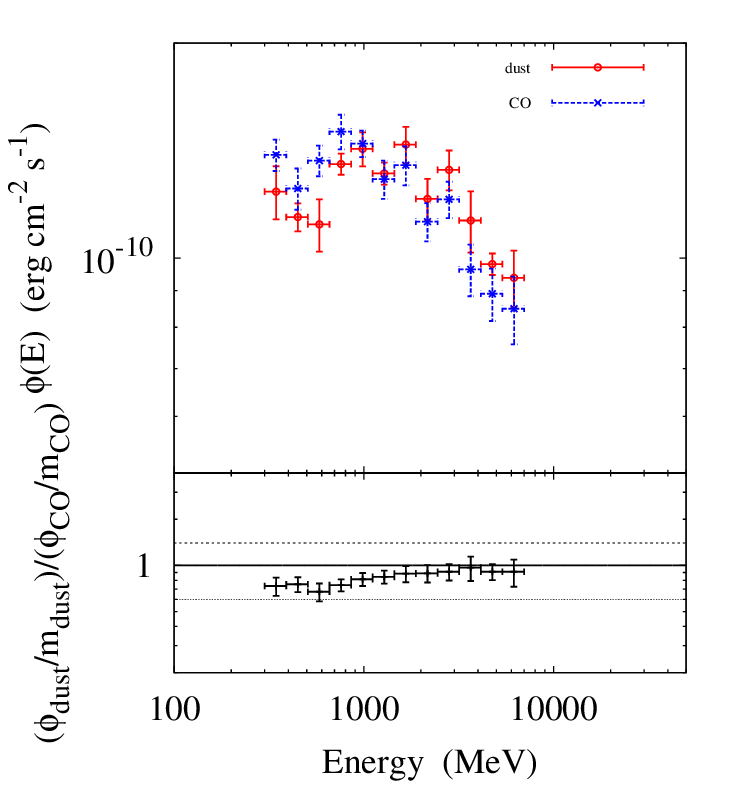}
\label{rhoOp:2}
}
\subfigure[][Perseus OB2]{
\includegraphics[scale=0.3]{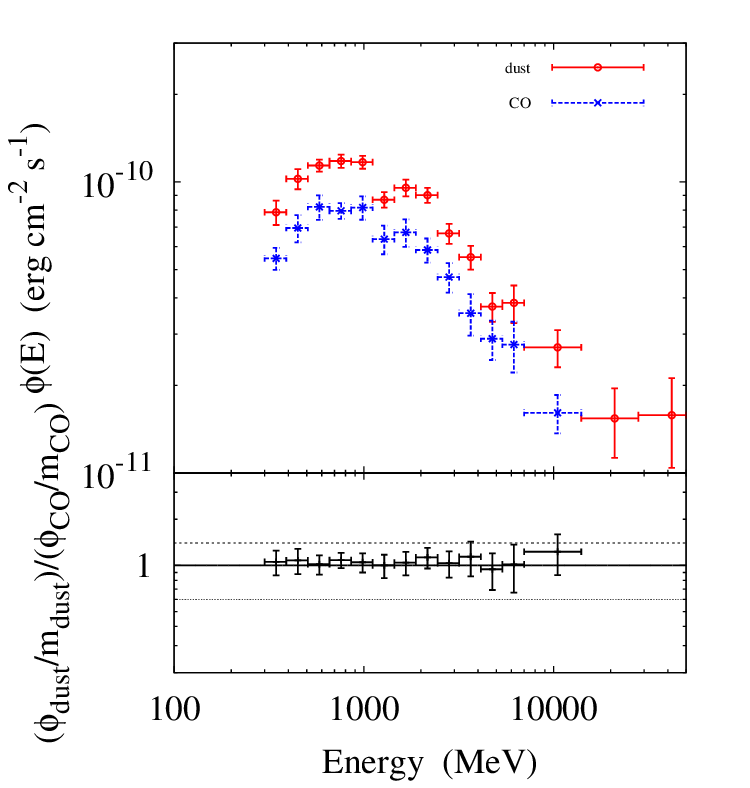}
\label{PerseusOB2:2}
} 
\subfigure[][Taurus]{
\includegraphics[scale=0.3]{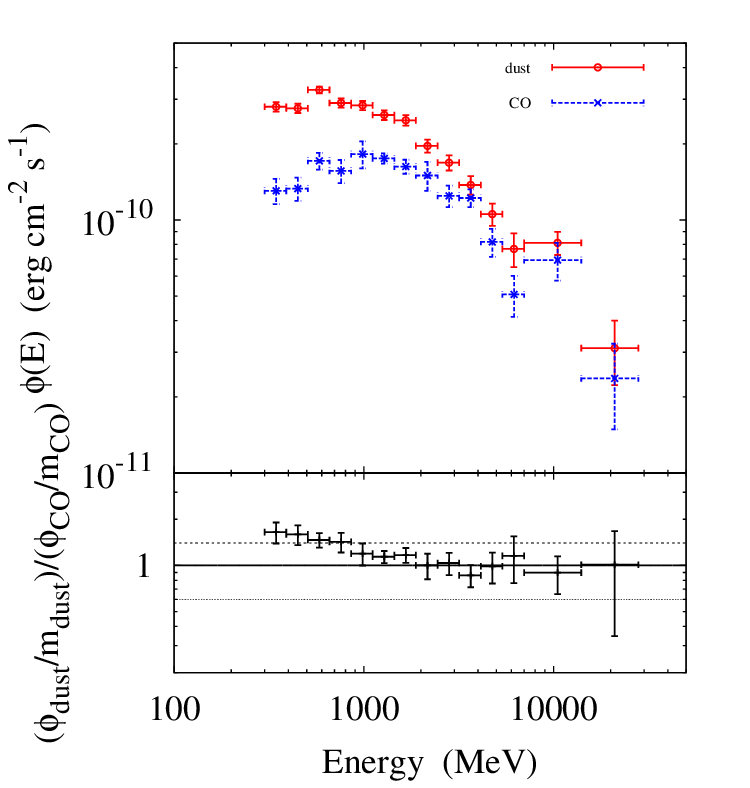}
\label{Taurus:2}
}
\subfigure[][Chamaeleon]{
\includegraphics[scale=0.3]{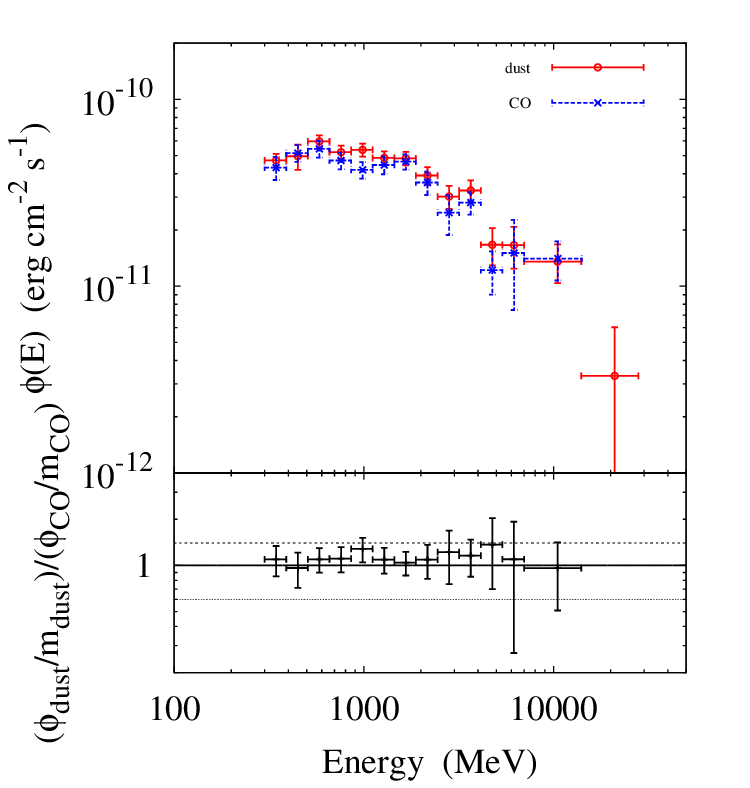}
\label{Chamaeleon:2}
}
\subfigure[][R CrA]{
\includegraphics[scale=0.3]{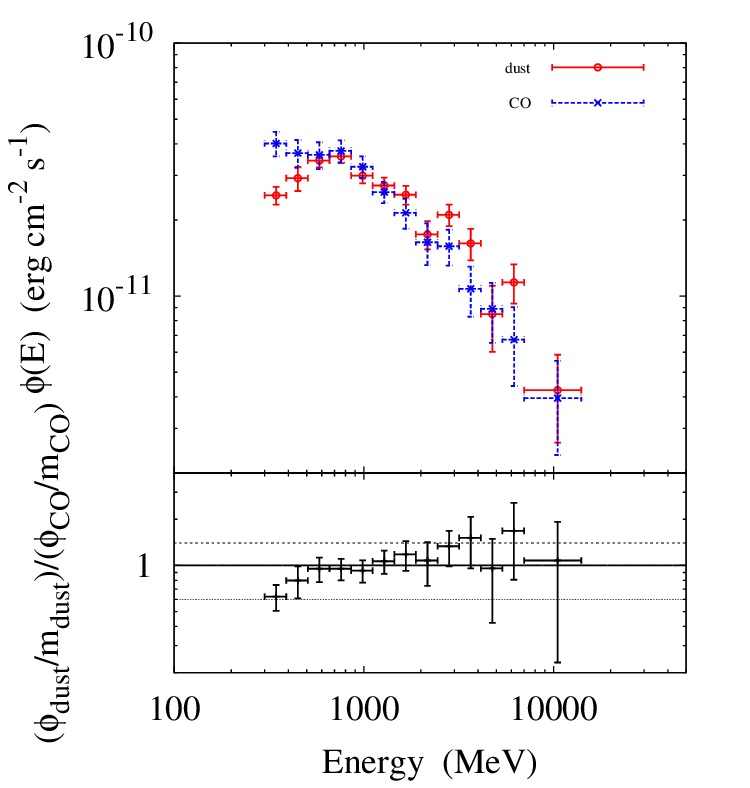}
\label{RCrA:2}
}
 \caption{SEDs derived from the templates generated from CO observation as well as from templates generated from Planck observation on dust opacity. The lower panel in every figure shows the ratio of $(\Psi_{dust}/M_{dust})/(\Psi_{CO}/M_{CO})$. The solid line in the lower panel marks when the ratio is equal to 1, whereas the dashed and dotted line show the uncertainties in deriving CO mass.} 
\label{fig:planck}
\end{figure*}

As a cross check we also use the templates generated in CO maps as the clouds' templates. In this case extra templates of emissions related to H~{\sc i}, calculated using GALPROP, are also included, since the CO map traces molecular gas only while dust opacity includes the contribution from both molecular and atomic gas.  The results for all eight clouds are shown in Figure \ref{fig:planck}.   It should be noted that the dust templates that consist of both H~{\sc i} and H$_{2}$ gas, as well as the possible dark gas, thus both the $\gamma$-ray flux and gas mass derived from the dust templates are higher than those from CO templates. However, the ratio flux/mass, which will determine the density of cosmic rays, are consistent in the two different choices of templates in most of the clouds. However, for one of them, Taurus, the spectrum shows some deviation, especially in the low energy range. This might be related to the fact that Taurus is much more diffuse, and the dust templates contain the regions where the CO density does not dominate, where the H~{\sc i} gas and dark gas cannot be ignored, which may have different $\gamma$-ray emissivities since their line-of-sight distance may be different from the CO gas.

 Recently, the Fermi collaboration has introduced  updated events reconstruction and instrument response functions, that has resulted in a significant increase in the effective detection area  (\cite{pass7}). This allows an extension of the analysis of the Fermi LAT dataset down to energy 100 MeV and below.  This is a crucial energy band to understand the origin of $\gamma-$ray emission. However,  although the effective area has been increased,  the uncertainties introduced by the diffuse background are still significant at low energies.  Therefore we present the analysis of the low energy part of Orion B, which has a high significance being detected ($TS \sim 4000$ above $300 \rm MeV$) and a low background level ($b<-10^{\circ}$ and far from the galactic centre). This is essential to minimize the uncertainties due to the diffuse background. The results  are shown in Appendix \ref{appendix2}.

\begin{figure*}[t!]
\centering
\subfigure[][{R CrA}]{
\includegraphics[width=0.25\linewidth]{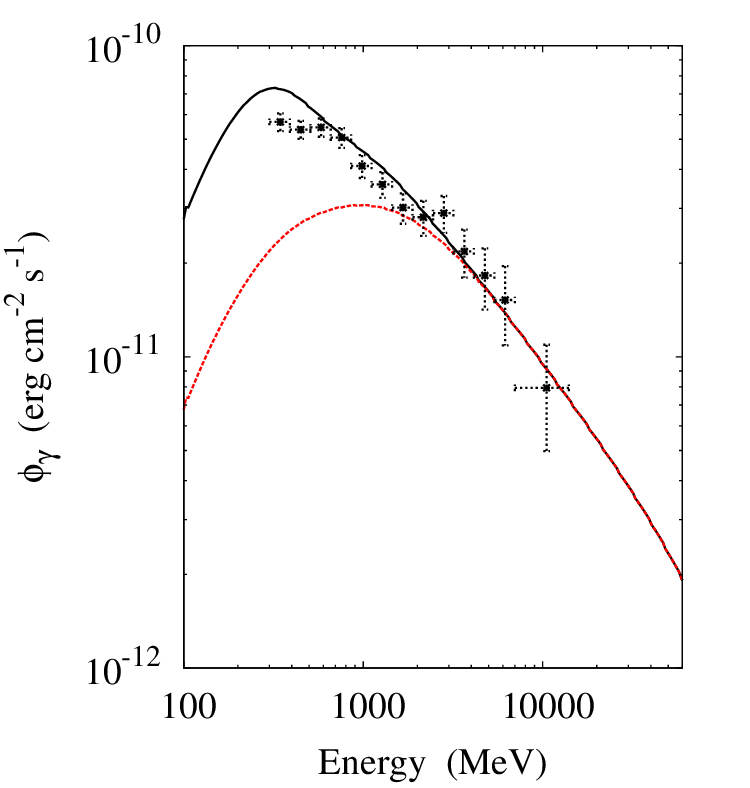}
\label{RCrA:1}
}
\subfigure[][{Orion B}]{
\includegraphics[width=0.25\linewidth,angle=0]{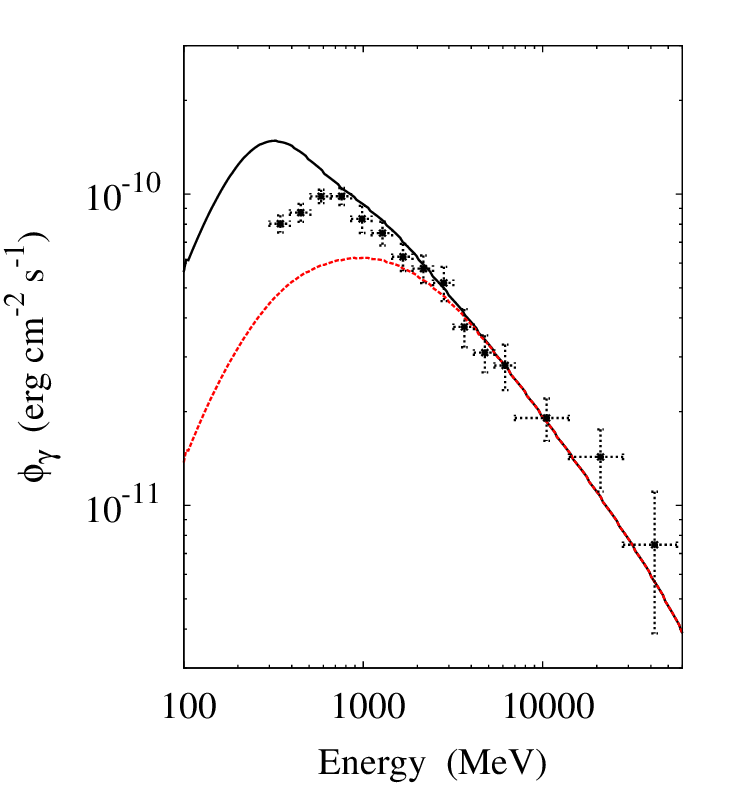}
\label{OrionB:1}
}
\subfigure[][{Perseus  OB2}]{
\includegraphics[width=0.25\linewidth]{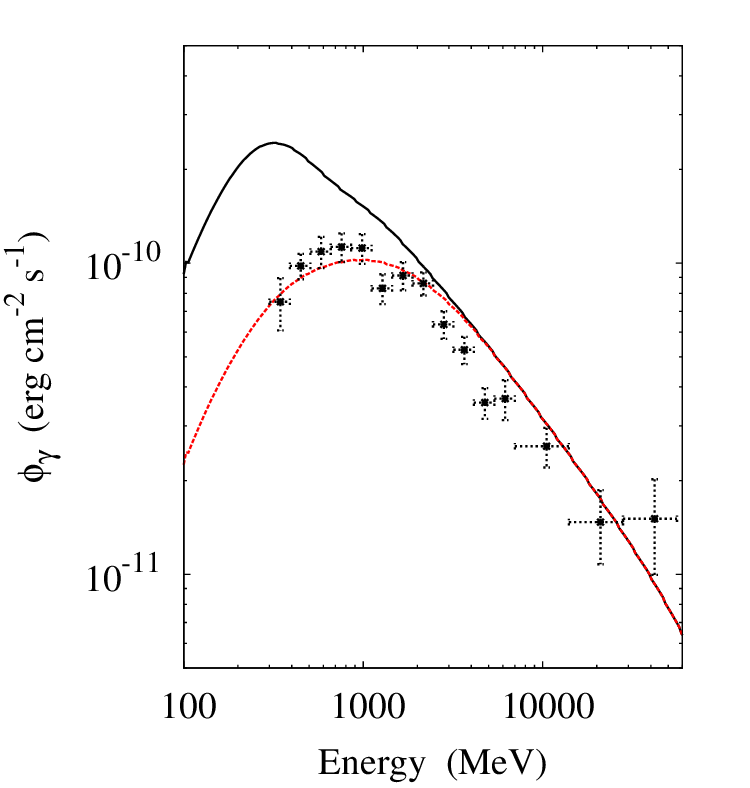}
\label{Per:1}
}

\caption{SED of three GMCs, R CrA (a),  Orion B (b)  and Perseus OB2 (c) as measured by LAT (points). For comparison, the expected $\gamma$-ray flux produced via pp interactions assuming that the CR spectrum inside the clouds is identical to the local one measured by PAMELA (red dashed lines) is also shown. The black solid curves represent the $\gamma$-ray fluxes calculated for a spectrum of CR protons which at high energies coincides with the PAMELA measurements, but towards low energies, down to 1~GeV, continues as a power-law with an index 2.85.}
\label{fig:3}
\end{figure*}

\section{Cosmic ray spectrum}
\label{section4}

The $\gamma$-ray spectra derived for each GMC from 300 MeV to about 30~GeV show different spectral shapes and flux levels (see Table 2 and Figure 1). Nevertheless the measured $\gamma$-ray flux level, provided that the total $\gamma$-ray emission in each cloud is due only to hadronic interactions of CR with the molecular target, is expected to be proportional to the ratio M/d$^2$. This general trend is shown in Figure \ref{fig:1}b, in which the differential flux at 3~GeV is plotted versus the M/d$^2$ ratio. The points are fitted with a simple linear regression that yields a
$\chi^2$/d.o.f of 3/9 (Prob=0.93). 

Owing to the high density of matter in these clouds, we assume that the $\gamma$-ray emission is produced mainly in interactions of galactic CRs and the gas inside the clouds. Under this hypothesis we can derive the CR spectrum and density inside each individual cloud.  To account for the contribution of the heavy nuclei we use the enhancement factor $1.85$ as suggested by \citet{mori09}. It should be noted that this value contains the enhancement both in the CR and in the interstellar medium. However, in calculating the mass of molecular clouds we have already included the contribution of heavy nuclei by using a factor of 1.36, thus the contribution from CR is represented by a factor of $1.85/1.36 \sim 1.36$ below. 

Recently a convenient formalism has been developed by \citet{vilante09} to derive proton spectrum directly from $\gamma$-ray observations. This method is based on the analytical parametrizations developed by \citet{kelner} for $\gamma$-ray spectra produced at the decays of neutral mesons, the secondary products of proton-proton (pp) interactions. While this parametrization provides good accuracy for protons with energy E$_{\rm  p}>$100~GeV, it shows some limitations at lower energies due to the approximation of the proton spectrum by a smooth power-law function.  To then derive the proton spectrum shape at low energy while avoiding an assumption of a power-law function to describe it, we chose instead the parametrization developed by \citet{kamae05}, which formally works for an arbitrary spectrum of protons in the energy range related to our data. It should be noted that this parametrization is less accurate at very high energies (E$_{\rm p}>$1~TeV, see e.g. Kachelriess \& Ostapchenko 2012), beyond the LAT sensitive energy range.

To investigate the $\gamma$-ray spectrum, especially at low energies, we compared the measured $\gamma$-ray flux in each cloud with the expected $\gamma$-ray emission assuming that the CR spectrum inside the clouds is identical to the one measured by the PAMELA experiment (local CR spectrum hereafter) \citep{pamela}. Such a comparison shows quite different behaviour for different clouds. In some of them, the spectrum is very similar to the one obtained when considering a CR spectrum with a low energy cutoff (likewise the local solar-modulated CR spectrum), whereas some others can be well-fit using a power-law function describing the CR spectrum down to energies  $1~\rm GeV$, as it is the generally assumed for the intrinsic spectrum of galactic CRs.

%\begin{landscape}
\begin{figure*}
\centering
\subfigure[][Orion A]{
\includegraphics[scale=0.33]{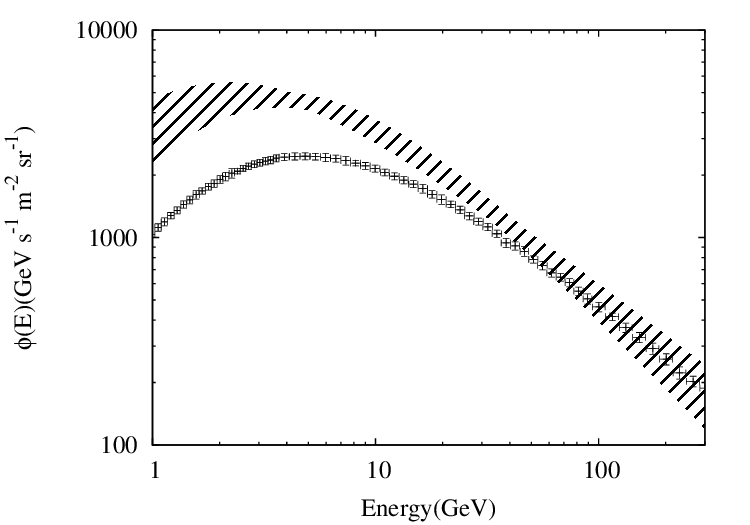}
\label{OrionA:3}
}
\subfigure[][Orion B]{
\includegraphics[scale=0.33]{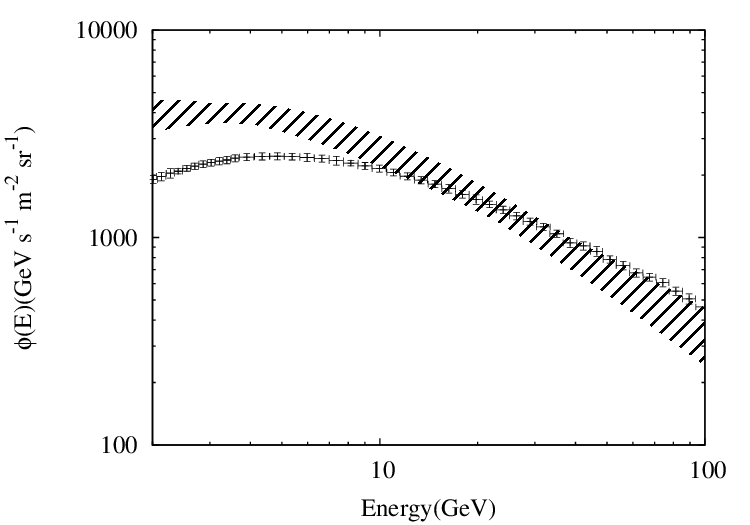}
\label{OrionB:3}
}
\subfigure[][Mon R2]{
\includegraphics[scale=0.33]{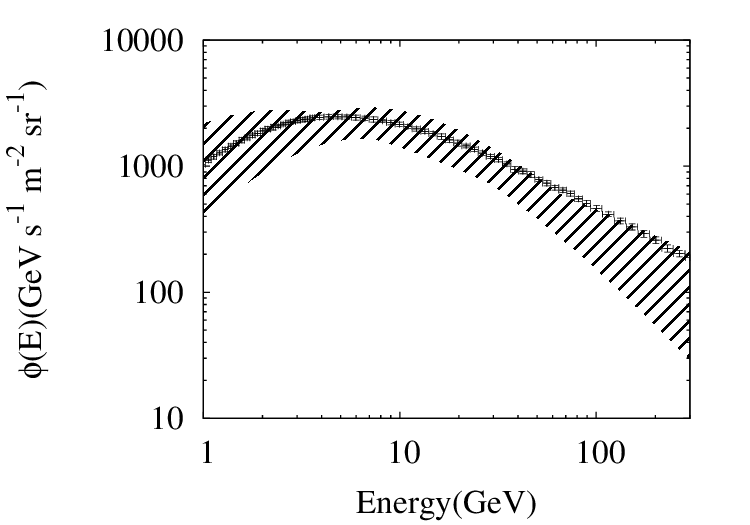}
\label{MonR2:3}
}
\subfigure[][$\rho$ Op]{
\includegraphics[scale=0.33]{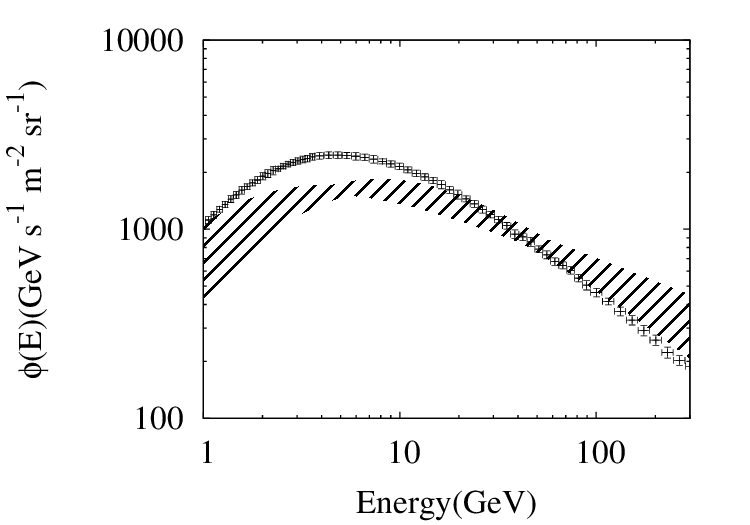}
\label{rhoOp:3}
}\\
\subfigure[][Perseus OB2]{
\includegraphics[scale=0.33]{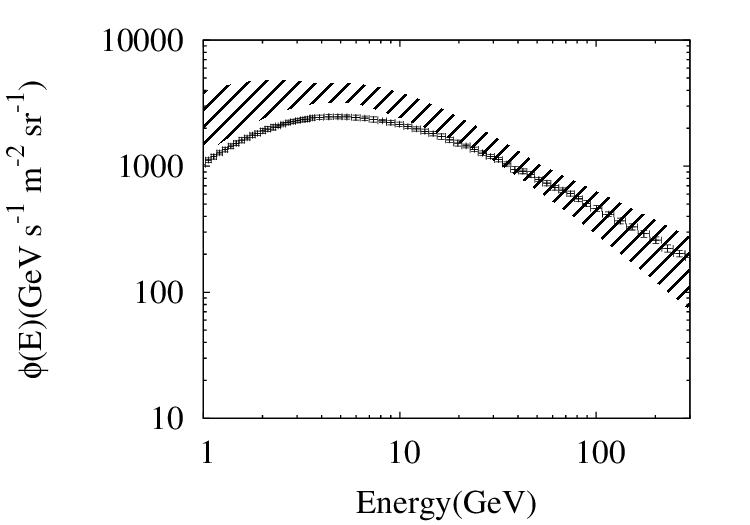}
\label{PerseusOB2:3}
} 
\subfigure[][Taurus]{
\includegraphics[scale=0.33]{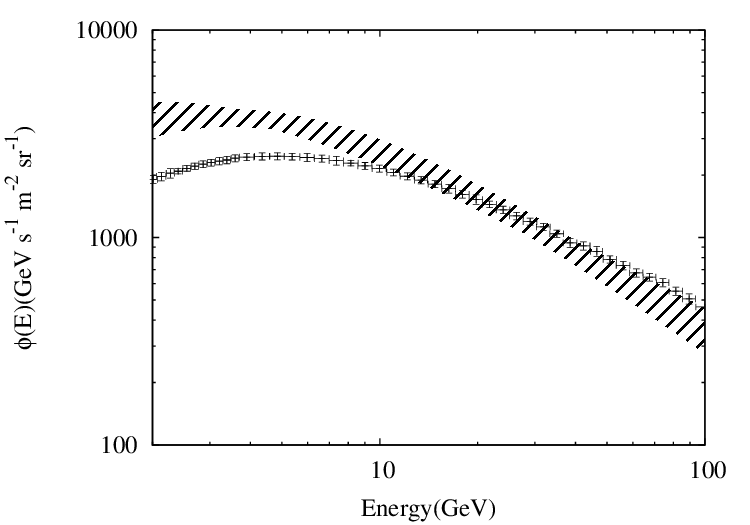}
\label{Taurus:3}
}
\subfigure[][Chamaeleon]{
\includegraphics[scale=0.33]{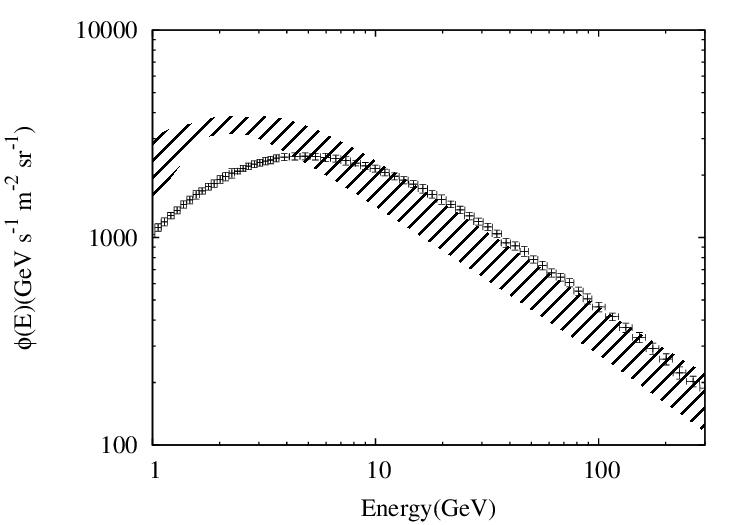}
\label{Chamaeleon:3}
}
\subfigure[][R CrA]{
\includegraphics[scale=0.33]{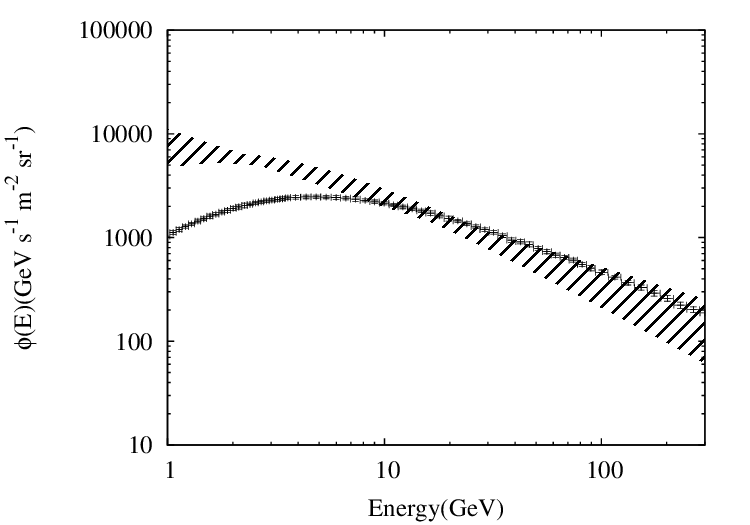}
\label{RCrA:3}
}
%\subfigure[][Aquila]{
%\includegraphics[scale=0.34]{aquila_br.eps}
%\label{Aquila:2}
%}
%\subfigure[][Hercules]{
%\includegraphics[scale=0.34]{hercules_br.eps}
%\label{Hercules:2}
%}
\caption{Energy spectra of CR protons in different clouds derived from the $\gamma$-ray data. It is assumed that 
the interactions of CR with the ambient gas are fully responsible for the observed $\gamma$-ray fluxes.
The shaded regions represent  1$\sigma$ fits for the proton  spectra. For comparison, the measurements 
of CR protons by PAMELA  are also shown (black  crosses).}
\label{fig:4}
\end{figure*}
\begin{figure*}
\centering
\subfigure[][Orion A]{
\includegraphics[scale=0.2]{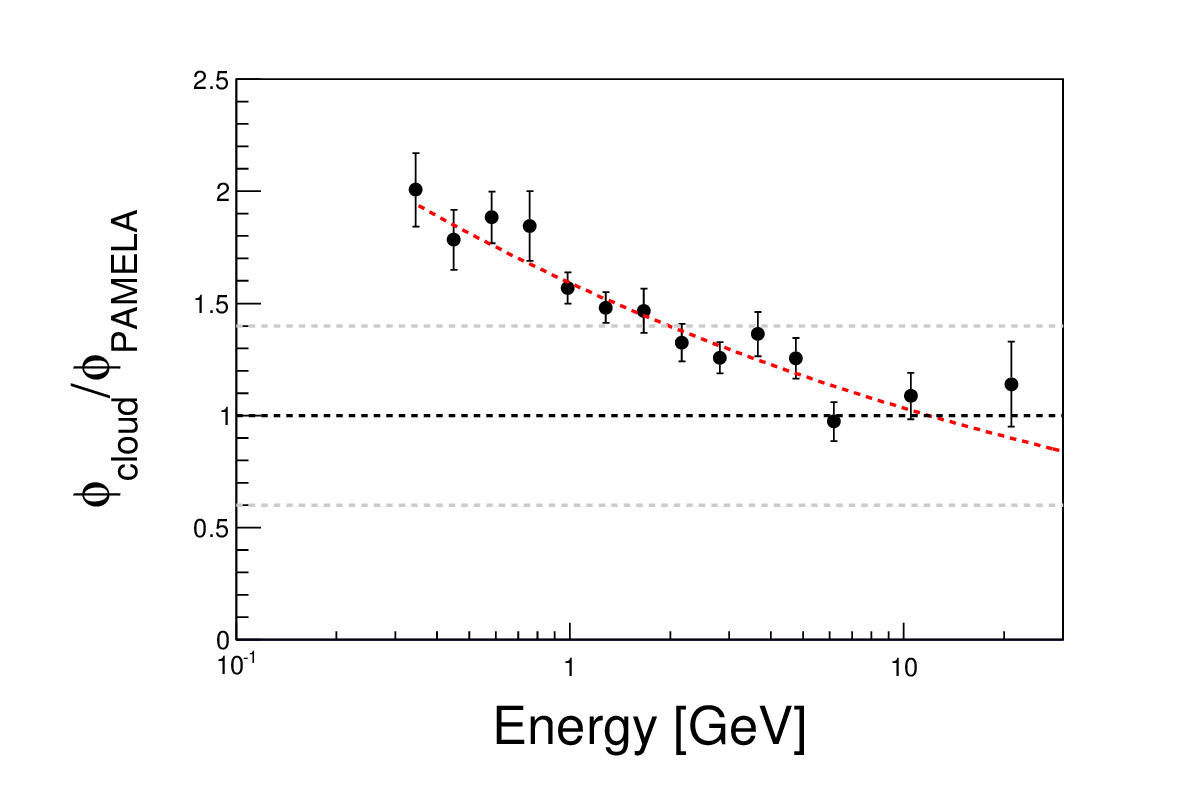}
\label{OrionA:4}
}
\subfigure[][Orion B]{
\includegraphics[scale=0.2]{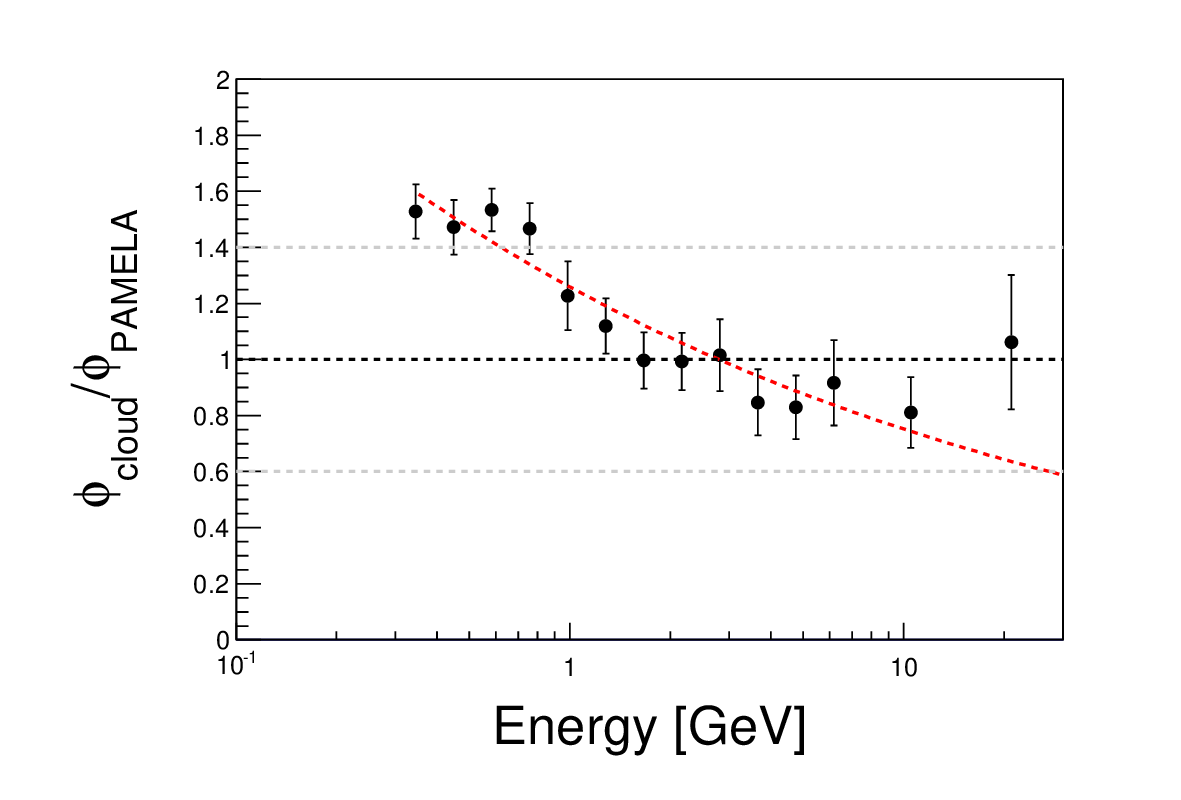}
\label{OrionB:4}
}
\subfigure[][Mon R2]{
\includegraphics[scale=0.2]{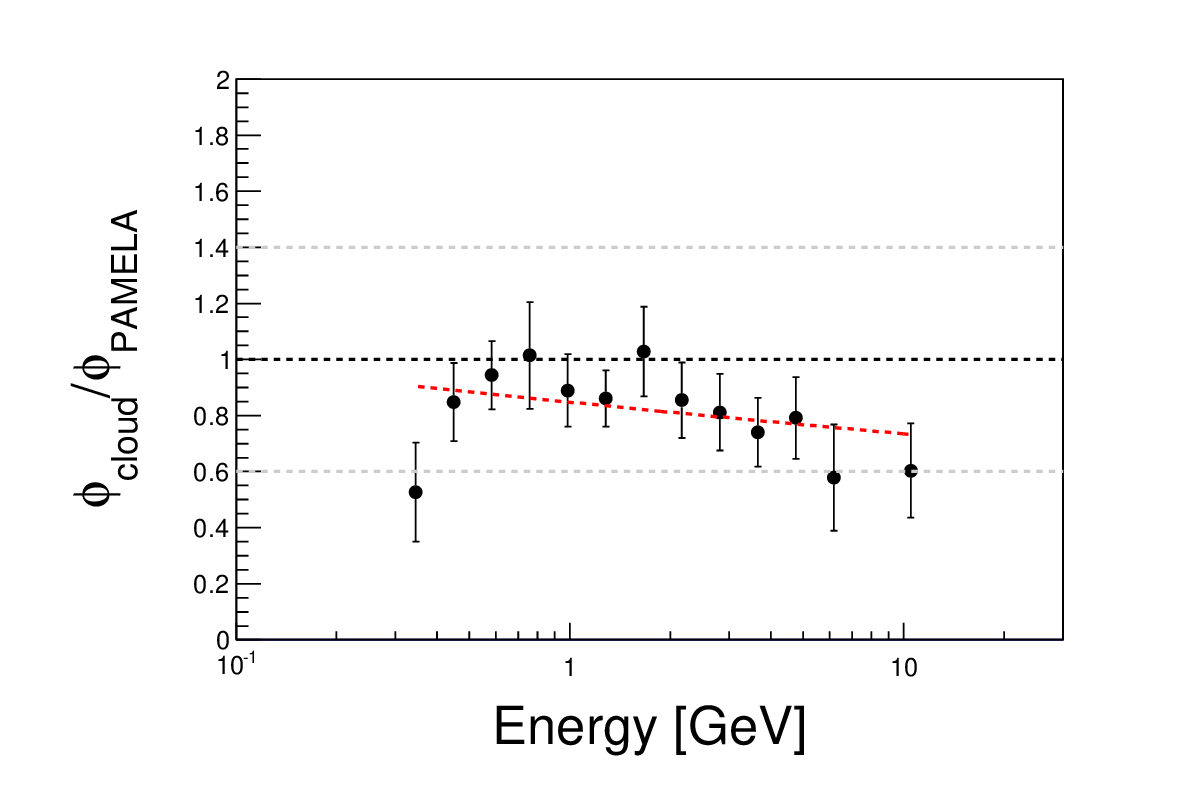}
\label{MonR2:4}
}
\subfigure[][$\rho$ Op]{
\includegraphics[scale=0.2]{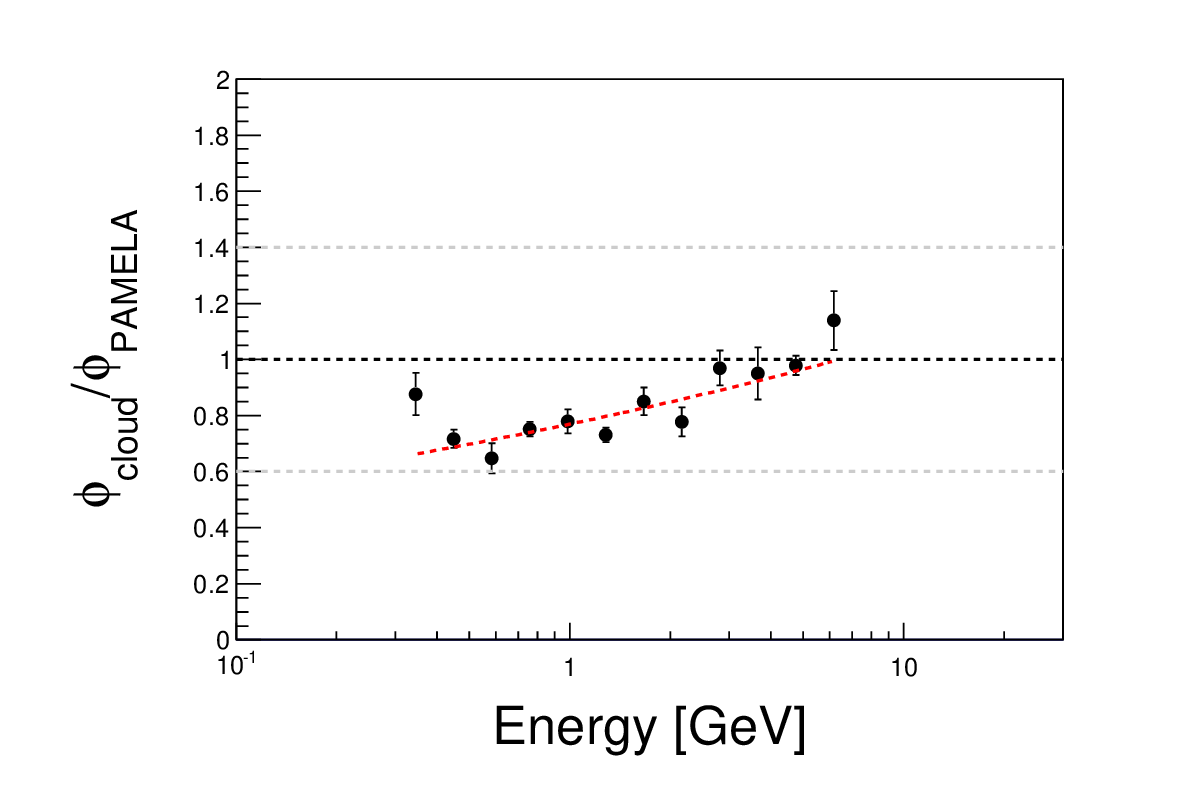}
\label{rhoOp:4}
}\\
\subfigure[][Perseus OB2]{
\includegraphics[scale=0.2]{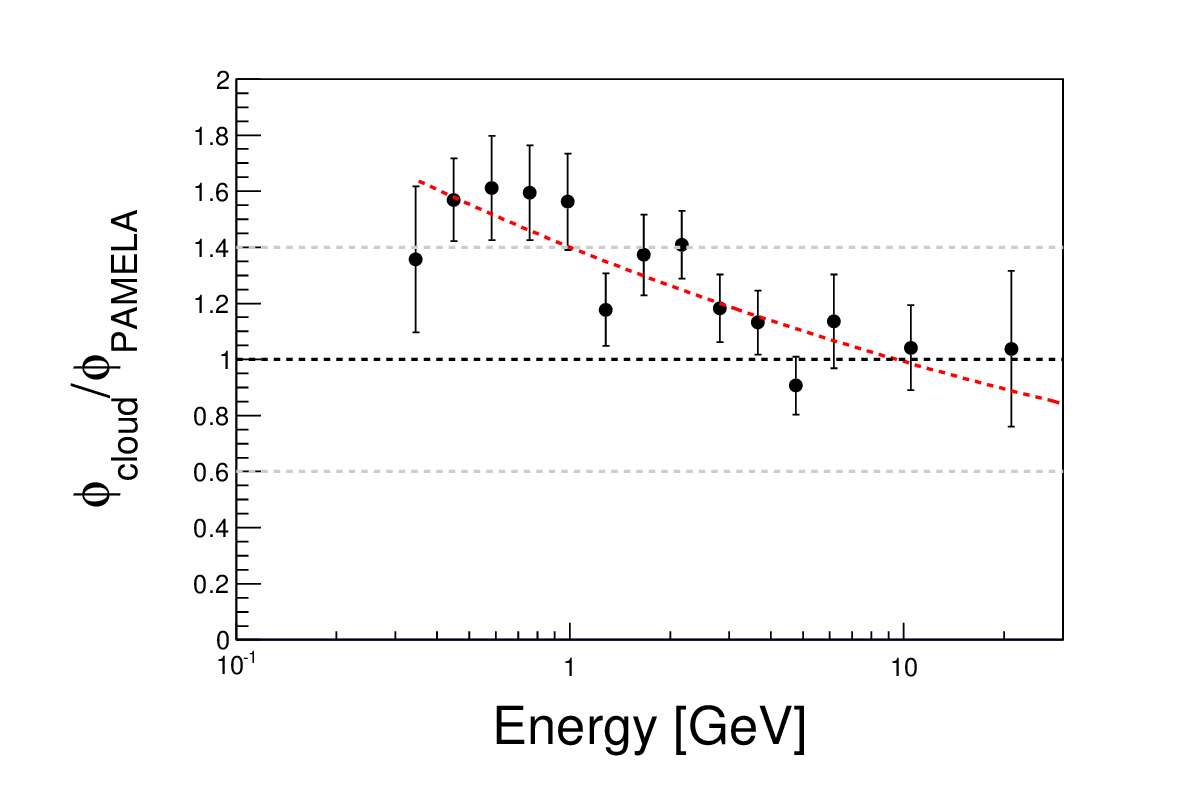}
\label{PerseusOB2:4}
} 
\subfigure[][Taurus]{
\includegraphics[scale=0.2]{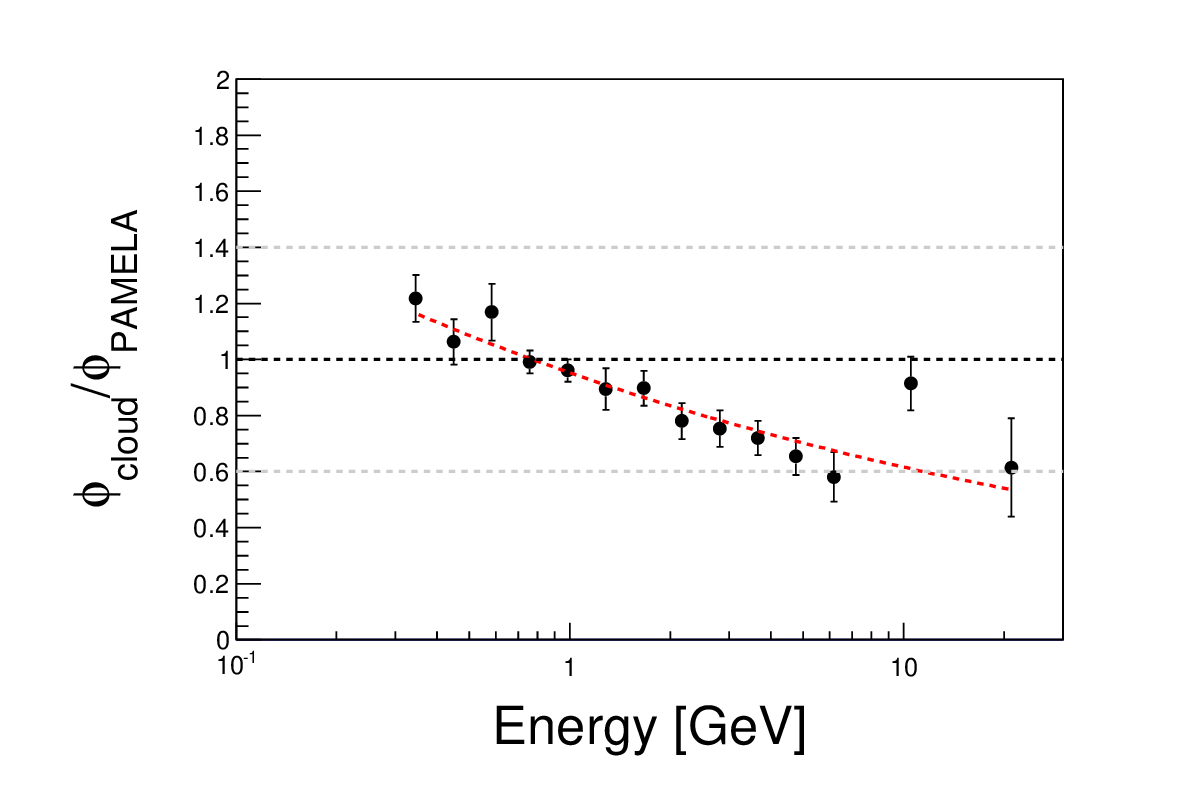}
\label{Taurus:4}
}
\subfigure[][Chamaeleon]{
\includegraphics[scale=0.2]{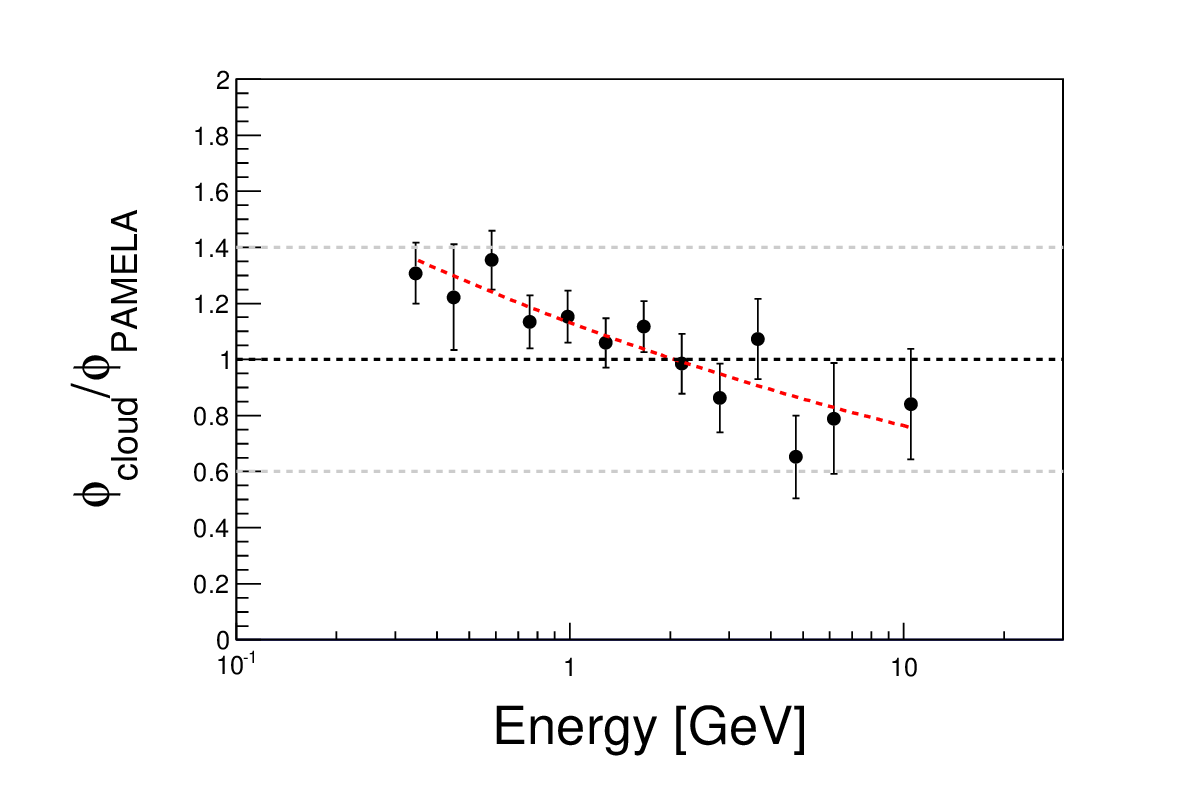}
\label{Chamaeleon:4}
}
\subfigure[][R CrA]{
\includegraphics[scale=0.2]{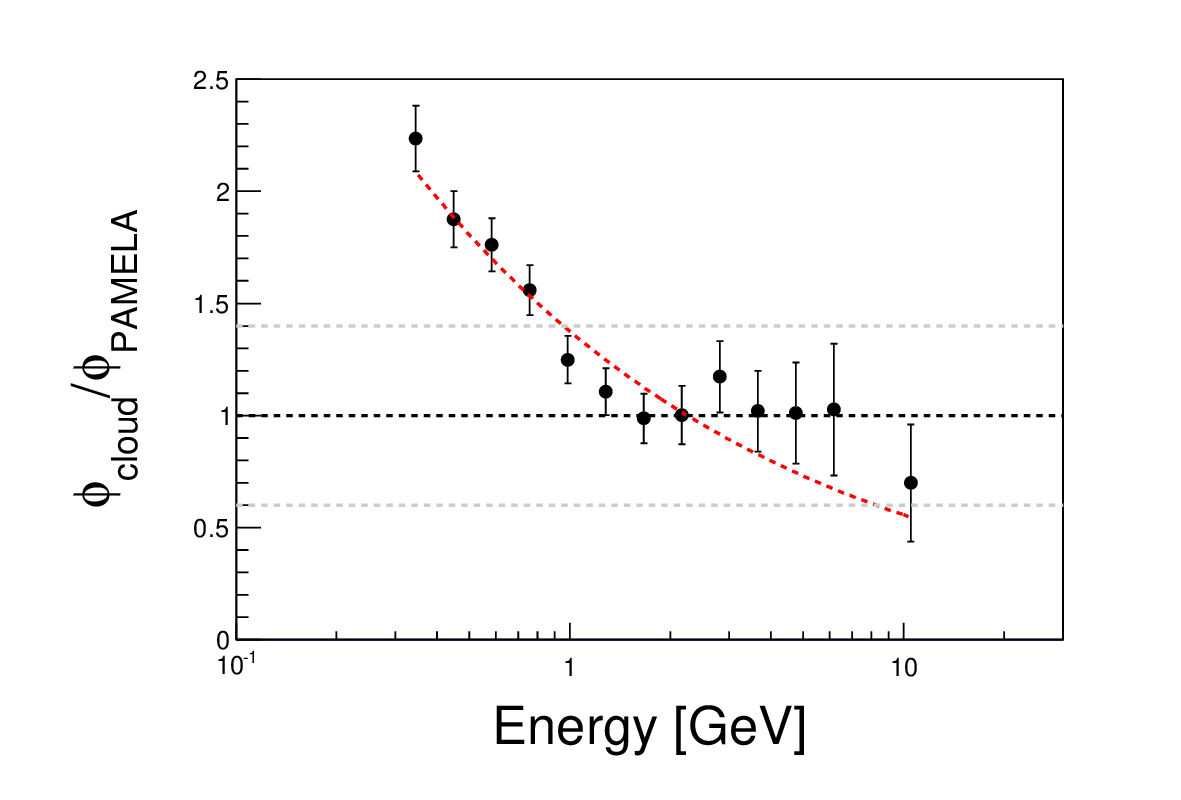}
\label{RCrA:4}
}
%\subfigure[][Aquila]{
%\includegraphics[scale=0.21]{aquila_ratio.eps}
%\label{Aquila:3}
%}
%\subfigure[][Hercules]{
%\includegraphics[scale=0.21]{hercules_ratio.eps}
%\label{Hercules:3}
%}
\caption{The ratio of $\gamma$-ray spectra  detected by Fermi LAT from different clouds to the 
calculated $\gamma$-ray spectrum if the CR proton flux  in the clouds are similar 
to the local CR spectrum measured by PAMELA. The grey lines  indicate the allowed range 
of variations related to uncertainties of the clouds' total masses.}
\label{fig:6}
\end{figure*}
%\end{landscape}

 Figure \ref{fig:3} shows three examples in which the measured spectral points of the $\gamma$-ray flux are overlaid with the expected $\gamma$-ray spectrum calculated using the local CR proton spectrum  and a pure power-law proton spectrum  normalized to PAMELA observations above 30 GeV (black solid line). While the observed spectrum of R CrA is described well by a $\gamma$-ray spectrum derived from a pure power-law CR spectrum, the two other sources shown, Orion B and Perseus OB2, show some deviation from a pure power-law. The spectrum measured in  Perseus OB2 agrees with the spectrum derived using the local solar-modulated CR spectrum, whereas the spectrum for Orion B seems to fall in between the two hypothesis considered, namely, with a harder spectrum at low energy than predicted by local solar-modulated CR spectrum but softer than the one predicted by a pure power-law spectrum.

To obtain the proton spectrum from the $\gamma$-ray observations, we used the statistical approach suggested by \citet{fit}. Three spectral shapes have been assumed for CR protons to fit the $\gamma$-ray data by the spectra of secondary ($\pi^0$-decay) $\gamma$ rays:
\noindent
(i) power law in kinetic energy  (KPL): 
\begin{equation} 
\psi(E)=N~E^{-\gamma} \ , 
\end{equation} 
\noindent
(ii) power law in total energy (TPL) 
\begin{equation}
\psi(E)=N~(E_{\rm total})^{-\gamma} \ ,
\end{equation}
\noindent
(iii) broken power law (BPL)
\begin{equation}
\psi(E)=N~(\frac{E}{E_b})^{-\gamma_1}[1+(E/E_b)^2]^\frac{\gamma_1-\gamma_2}{2} \ , 
\end{equation}
where $E$ and $E_{\rm total}=E+m_{\rm p} c^2$ are the proton kinetic and total energies, respectively; $E_{\rm b}$ corresponds to the break energy,  where the index of the power law distribution smoothly changes from $\gamma_1$  to $\gamma_2$.

\begin{figure}
\centering
\includegraphics[width=0.8\linewidth]{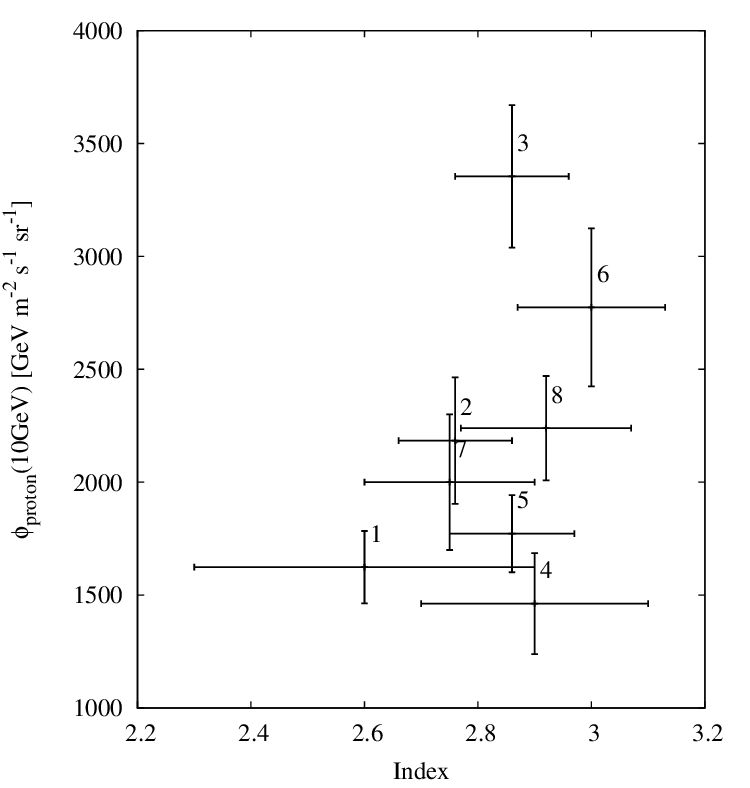}
\caption{The power-law  index of CR protons above $E_b$ versus differential proton flux at $10~\rm GeV$. The data points are numbered as in Table 2.}
\label{fig:5}
\end{figure}

The $\chi ^2/ d.o.f$ values and associated probabilities for fitting each spectral template are listed in Table \ref{tab:2}.  Orion B, Hercules, $\rho$ Oph and R CrA are fitted well by the three representations at the 2$\sigma$ confidence level. In the other cases, the KPL model is disfavoured by more than 2.5 $\sigma$ whereas the difference between the TPL and BPL models is not significant. Figure \ref{fig:4} shows the resulted proton spectrum using the BPL hypothesis, where the shaded area accounts for the 1$\sigma$ error in the fit parameter space. For
comparison, the local proton spectrum measured by PAMELA is also shown (crosses).

From the comparison with the local proton spectrum we can conclude:
 \begin{itemize}
\item The proton spectrum derived from three out of the eight clouds considered, namely Taurus, Persues OB2, and Mon R2  are described well by the locally measured proton spectrum by PAMELA.
%\item The derived proton spectra of Aquila  shows a deficit at high energies but it should be noted in this regard that this object is located in the direction of the galactic plane, where the uncertainties in the background model do not allow accurate derivation of the $\gamma$-ray spectrum. 
\item For $\rho$ Oph, the derived cosmic ray spectrum is harder and total flux is lower than those observed by PAMELA ($2.6 \pm 0.3$).  However,   as mentioned above, this source is overlaid with several bright point sources, and the SED of $\rho$~Oph only extend to $\approx$7\,GeV, preventing a good determination of the cosmic ray flux and spectrum at high energies.

\item The four other cases, Orion A, Orion B , Chamaeleon, and R CrA should be treated differently. In R CrA the proton spectrum is
  described well with a KPL function, showing an enhancement of CRs compared to the local CR flux.  In Orion A, Orion B, and Chamaeleon, although a BPL function is preferred to describe the proton spectrum, a high enhancement of CRs compared to the local CR flux at low energies is also present. At high energies the spectrum shape is very similar to the one of the local CR. It is important to note that R CrA is the only GMC considered which is not located inside the Gould Belt, implying a common behaviour at high-energies independently of the cloud location.
\end{itemize}

Figure \ref{fig:5} shows the distribution of the CR spectrum index at high energy ($E > E_{\rm b}$) with respect to the CR flux at 10~GeV. Most of the spectral indices found are compatible ($<$2$\sigma$) with the CR \emph{sea} spectral index of 2.8. %The only exception is the one corresponding to Aquila (3$\sigma$ deviation) which suffers from background subtraction as discussed above. 
The index for $\rho$~Oph is also a little smaller ($2.6 \pm 0.3$) than 2.8. However,  as mentioned above, the SED of $\rho$~Oph only extends to $\approx$7\,GeV and may be also influenced by several bright point sources which are overlaid with the cloud, preventing a good determination of the spectral index at high energies.

To further explore the impact of deviation of the proton spectra in the clouds from the locally measured CR flux, we calculated the ratio of the measured $\gamma$-ray flux to the expected one for each cloud under the hypothesis that the CR flux inside the cloud is identical to the local CR flux. These results are shown in Figure \ref{fig:6}. The indicated error bars account for both
statistic and systematic errors. The effects discussed above can also be observed here: the SED of Orion B, Orion A, Chamaeleon and R CrA shows different
spectral features compared to the expected $\gamma$-ray spectrum produced by the local CRs at low energies. The same effect is observed in R CrA (not associated with the Gould Belt) where the CR spectrum shows a clear deviation from the predicted one.

\section{Summary}

Giant  molecular clouds can serve as  unique \emph{barometers} for determining the pressure (energy density) of CRs throughout the Galaxy via their characteristic high energy $\gamma$-ray emission produced in CR interactions with the dense ambient  gas. Similar information can also be found by analyzing the diffuse $\gamma$-ray background of the galactic disk  in different directions. However in the case of the diffuse $\gamma$-ray background, the extracted CR spectrum is averaged over huge distances on a $\geq 10$~kpc scale, while the information about CRs derived from individual clouds is localized within tens of parsecs. 
  
The high quality data obtained by Fermi LAT provide adequate observational material for such studies. In this paper we present the result of our analysis of a five-year observation of eight GMCs detected by Fermi LAT. At high energies, the analysis of $\gamma$-ray emission from all GMCs, except for $\rho $ Oph, which have limited statistics, show that the CR proton spectra above a few GeV are described well by a power-law function with an index of $\Gamma=2.85$. This is slightly softer than previous measurements (e.g. Boezio et al. 1999) but agrees very well with the CR proton spectral index reported by the PAMELA collaboration \citep{pamela}. Remarkably, the derived absolute fluxes of CRs  also agree with the direct measurements of local CRs. The conclusion on the absolute fluxes depends highly on the mass cloud estimation. Here we compare the masses derived from CO radio observations and dust, obtaining compatible results. This enforces the reliability of the absolute CR fluxes we measured in the clouds, at the level of the one measured by PAMELA.

%It should be noted, however, that this conclusion is based on the estimates of the clouds mass based on CO observations which contain rather large, up to 50 \% uncertainties. Moreover, it has been recently argued (see i.e. \citealt{ade} and references therein) that the gas is not adequately traced by CO measurements, and the total mass could  be underestimated up to by a factor of two. Whether this claim concerns the specific clouds used in this paper, and if yes, then to which extent, this is a question which will be clarified, hopefully soon. Meanwhile, we should worn that any substantial revision of masses of the clouds in the Gould Belt, which could be only towards the increase of the mass estimates would unavoidably lead to a surprising conclusion, namely it would lead to the reduction of the estimates of the CR energy flux well below the level locally measured by Pamela. This seems rather unlikely, but cannot be fully excluded. The prime importance of this question for the origin of galactic CRs apparently initiate new model-independent measurements of masses of clouds in Gould Belt.

On Earth, at low energies, typically below 10 GeV, we should expect a strong suppression of the flux of local CRs due to modulation during propagation through the solar system. This effect can be seen in the spectrum reported by the Pamela collaboration. The \emph{recovery} of the initial (unmodulated) CR spectrum at low energies is a very difficult task and depends on theoretical (model) assumptions. This increases the importance of deriving the spectra of galactic CRs from $\gamma$-ray observations of \emph{passive} clouds. However, propagation effects might have a strong impact on the penetration of low energy CRs into the GMCs, and therefore also on the spectrum of secondary $\gamma$ rays. The $\gamma$-ray spectra of six out of eight GMCs used in the present study (see Figure \ref{fig:5}), do not significantly deviate from the expected one if the CR spectra in these clouds were similar to the locally measured CR spectrum. In this regard, it should be noted that the suppression of the spectrum of local CRs is unavoidable due to the solar modulation \citep{pamela2}. 

Generally, we should expect non-negligible modulation of the spectrum of galactic CRs at their entrance into the complexes of dense molecular clouds and star formation regions as well.  However, we suspect that the modulation of galactic CRs in such different environments (the solar system and massive molecular clouds) could have the same effect on the deformation of the spectrum of galactic CRs. At first glance, a possible explanation of this coincidence could be that the modulation effects are negligible and the local CRs spectrum measured by PAMELA is the same as the galactic CR spectrum not only at high energies but also below 10~GeV.  However, since there could be little doubt as to the significant modulation of CRs in the solar system, it is more likely that the similarity of the CR spectra in clouds and the local CR spectrum measured by PAMELA at low energies does not have a deep physical reason, especially given that only in two cases, $\rho$~Oph and Taurus, the CR proton spectra mimic the PAMELA spectrum. Moreover, the good representation of the $\gamma$-ray spectra of Orion~A,  Orion~B, Chamaeleon, and  R~CrA (see Figure \ref{fig:3}) by interactions of CRs with pure power-law spectra of parent protons down to 1 to 3 GeV, supports the interpretation of the diversity of spectral shapes of CRs in different clouds below 10~GeV as a result of propagation effect. In this regard, the $\gamma$-ray spectra of these four clouds indicate a weak effect of particle propagation in these clouds, therefore the power-law spectra of protons derived in these  clouds could be taken as representative of the \emph{sea} of galactic CRs. The energy density corresponding to this spectrum (which continuous as a power law in kinetic energy with index $\Gamma \approx 2.85$ from energies $\geq 100$~GeV down to 2 GeV) is $\sim$ $0.65 ~  \rm eV/cm^3$, while  for the local CR spectrum measured by PAMELA the density is a factor of 4 less. 

Finally, combining this spectrum with the well established (in this energy interval) energy-dependent time of escape of CRs from the Galaxy, $\tau(E) \propto  E^{-\delta}$ with $\delta \approx 0.5-0.6$, implies a source (acceleration) of galactic CRs below 100~GeV with a spectral index close to $E^{-2.3}$, which is somewhat steeper than one would expect from the standard diffusive shock acceleration theory.  

The determination of the low energy galactic CR spectrum has important implications on various astrophysical issues, such as ionisation and heating of ISM by low energy CRs and contribution to the overall pressure in the ISM. It also has an impact on the determination of the diffuse $\gamma$-ray background, which is crucial for correctly interpreting of the high energy observations. Further studies of processes involving propagation and $\gamma$-ray production by CRs in massive molecular clouds are needed. In particular, the continuation of the analysis of the Fermi LAT data based on more time exposure of these objects, is important for the search for smaller differences in the $\gamma$-ray spectra.

\appendix

 \section{Gamma-ray production mechanisms in molecular clouds}
\label{apendix1}
$\gamma$ rays are produced in molecular clouds through different radiation mechanisms, namely in proton-proton collision via production and decay of mainly neutral pions, due to electron and positron bremsstrahlung, as well as inverse Compton (IC) scattering of  the primary and secondary electrons. To show the contributions of different mechanisms, we consider a molecular cloud with a mass of $10^5 M_{\odot}$ located $1 ~\rm kpc$ away from us. The proton and primary electron spectra measured by PAMELA \citep{pamela,pamelae} are used to calculate the gamma ray spectrum. We show the $\gamma$-ray spectra for CR distributions extrapolated from high energies as a pure power law down to low energies.

For the spectrum of secondary electrons, we take the radiative energy losses into account, while assuming that electrons do not escape the cloud. This implies that the corresponding fluxes of the secondary $\gamma$ rays should be treated as upper limits. Fig\ref{fig:gamma} shows that at all energies above 100~MeV the pion decay dominates the contributions from the electrons. While the IC fluxes are negligible, around 100 MeV the $\gamma$-ray flux from bremsstrahlung of primary electrons becomes significant, approximately $25\%$ (of the $\pi^0$-decay $\gamma$-ray flux) for the locally measured electron spectrum and almost $100\%$ for the CR spectrum extrapolated from high energies as a pure power law down to low energies. The electron spectrum in the interstellar medium below 1 GeV derived from the radio synchrotron measurements is between these two approximations \citep{strong11}, therefore the curves shown in Figure \ref{fig:gamma} can be considered as the lower and upper limits of the contribution of bremsstrahlung of primary electrons.

The bremsstrahlung of secondary electrons may be important as well (up to  about $50\%$ at 100MeV); however, in the case of effective escape of these particles from the cloud, their contribution would be dramatically reduced. Finally we note the difference of $\gamma$-ray spectra produced at energies below a few GeV by the PAMELA-type and pure power-law proton spectra. Since the spectrum measured by PAMELA is strongly modulated  in the solar system, the proton spectrum with power-law shape extending down to low energies seems more realistic for the galactic CRs. Nevertheless, if the entrance of low-energy galactic CRs into the cloud is prevented, the $\gamma$-ray fluxes at low energies will be suppressed. Thus, the two curves corresponding to $\pi^0$ $\gamma$ rays in Figure \ref{fig:gamma} should be considered as lower and upper limits  if we neglect local accelerations.   
 
\section{Low energy $\gamma$-ray emission of Orion B}
\label{appendix2}

\begin{figure}
\centering
\includegraphics[width=0.8\linewidth]{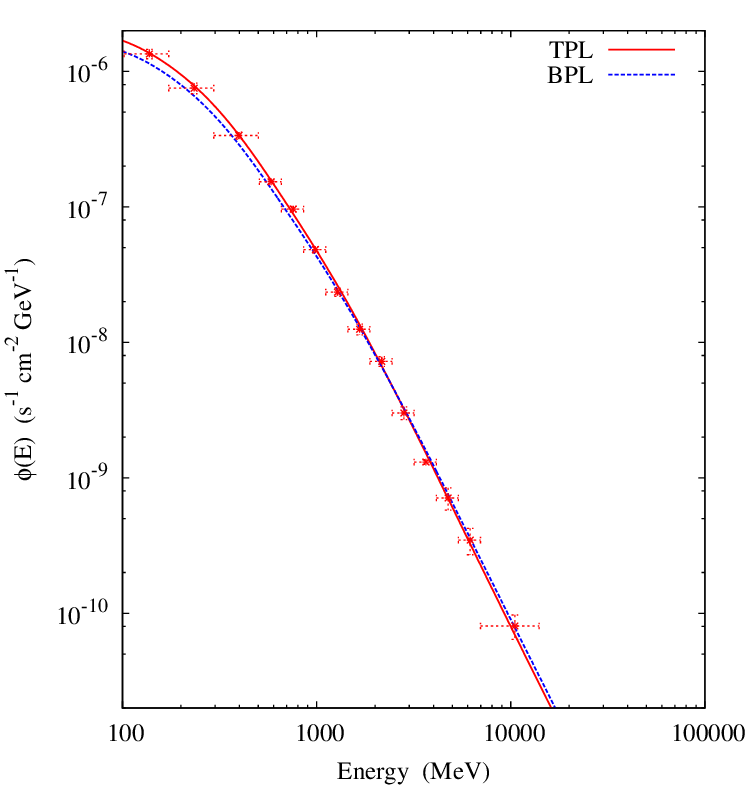}
\caption{The differential spectrum for Orion B including the low energy data points down to 100 MeV.  The dotted curve was calculated using the BPL proton spectrum described in the text with $E_b = 2~\rm GeV, \gamma_1=1.0$ and $\gamma_2=2.84$, while the solid curve using TPL model with  $\gamma=2.88$ }.
\label{fig:le}
\end{figure}

\begin{figure}
\centering
\includegraphics[width=0.8\linewidth]{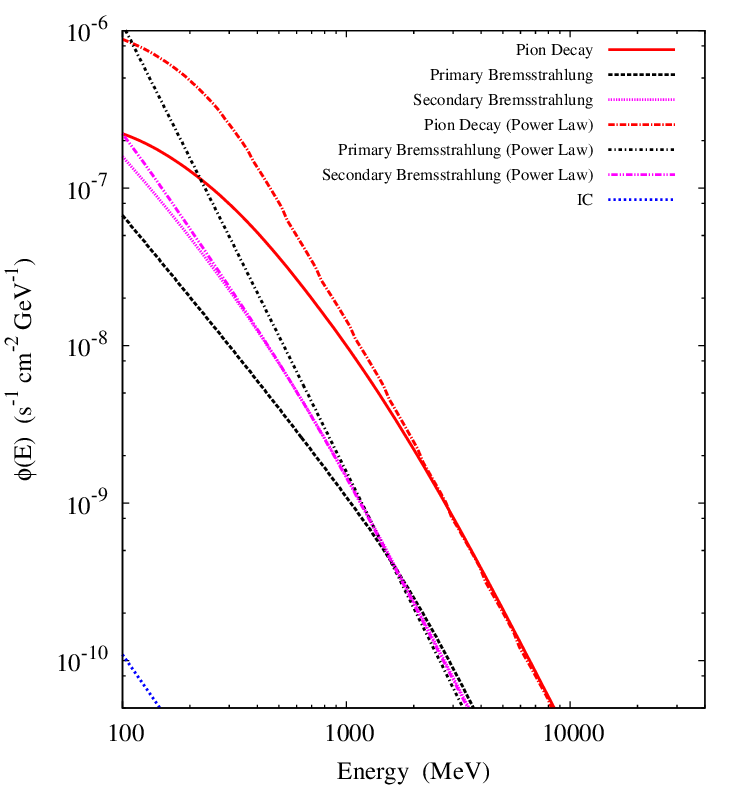}
\caption{The differential spectrum of $\gamma$-rays produced in molecular clouds by different mechanisms. Both the PAMELA measurement and a pure power-law function extrapolated to low energies  are considered.}
\label{fig:gamma}
\end{figure}

To derive the SED down to 100 MeV from Orion B, we use the same analysis method as described in the text, modifying the cut in energy to include low-energy photons. The derived $\gamma$-ray flux is shown in Figure \ref{fig:le}. The differential spectrum in the low energy bins shows a flattening that most likely corresponds to the distinct  feature expected in the  pion-decay spectrum. This is demonstrated by the theoretical spectra of $\gamma$ rays which are also shown in Figure \ref{fig:le}. These curves are obtained for two types of proton spectra:  BPL and TPL (as  described in the text). For the BPL hypothesis we adopt $E_b = 2~\rm GeV, \gamma_1=1.0$ and $\gamma_2=2.84$ while for the TPL model the index was set to be $\gamma=2.88$. These indices are close to the one observed by PAMELA \citep{pamela}. The results are shown in solid and dotted curves in Figure \ref{fig:le}, respectively. We would like to point out the $\gamma$-ray fluxes in the entire energy band, including the low energy part, can be  fitted quite well by pure $\pi^0$ contributions. Also we note that there is not any need to assume a break in the CR spectrum down to 1-2 GeV.  However, the contribution of the primary and secondary electrons which at 100 MeV can be as large as 50 \%. If so, a break in the proton spectrum at energy somewhat higher than 2~GeV might be needed.  
Also, at low energies the uncertainty in the diffuse background could be quite large, and this should be investigated carefully before we draw any conclusion concerning the proton spectrum around a few GeV.  

\bibliographystyle{aa}
\bibliography{mc_lan}

\begin{thebibliography}{30}
\expandafter\ifx\csname natexlab\endcsname\relax\def\natexlab#1{#1}\fi

\bibitem[{Abdo {et~al.}(2012)}]{2fgl}
Abdo {et~al.} 2012, Astrophys.J.Suppl., 199, 31

\bibitem[{Ackermann {et~al.}(2012a)Ackermann, Ajello, Allafort, Antolini,
  Baldini, {et~al.}}]{fermiorion}
Ackermann, M., Ajello, M., Allafort, A., {et~al.} 2012a, Astrophys.J., 756, 4

\bibitem[{{Ackermann} {et~al.}(2012b){Ackermann}, {Ajello}, {Allafort},
  {Baldini}, {Ballet}, {Barbiellini}, {Bastieri}, {Bechtol}, {Bellazzini},
  {Berenji}, {Blandford}, {Bloom}, {Bonamente}, {Borgland}, {Bottacini},
  {Brandt}, {Bregeon}, {Brigida}, {Bruel}, {Buehler}, {Busetto}, {Buson},
  {Caliandro}, {Cameron}, {Caraveo}, {Casandjian}, {Cecchi}, {Charles},
  {Chekhtman}, {Chiang}, {Ciprini}, {Claus}, {Cohen-Tanugi}, {Conrad},
  {D'Ammando}, {de Angelis}, {de Palma}, {Dermer}, {Digel}, {Silva}, {Drell},
  {Drlica-Wagner}, {Falletti}, {Favuzzi}, {Fegan}, {Ferrara}, {Focke},
  {Fukazawa}, {Fukui}, {Funk}, {Fusco}, {Gargano}, {Gasparrini}, {Germani},
  {Giglietto}, {Giordano}, {Giroletti}, {Glanzman}, {Godfrey}, {Grenier},
  {Grondin}, {Grove}, {Guiriec}, {Hadasch}, {Hanabata}, {Harding}, {Hayashi},
  {Horan}, {Hou}, {Hughes}, {Itoh}, {Jackson}, {J{\'o}hannesson}, {Johnson},
  {Kamae}, {Katagiri}, {Kataoka}, {Kn{\"o}dlseder}, {Kuss}, {Lande}, {Larsson},
  {Lee}, {Lemoine-Goumard}, {Longo}, {Loparco}, {Lovellette}, {Lubrano},
  {Martin}, {Mazziotta}, {McEnery}, {Mehault}, {Michelson}, {Mitthumsiri},
  {Mizuno}, {Moiseev}, {Monte}, {Monzani}, {Morselli}, {Moskalenko}, {Murgia},
  {Naumann-Godo}, {Nemmen}, {Nishino}, {Norris}, {Nuss}, {Ohno}, {Ohsugi},
  {Okumura}, {Omodei}, {Orlando}, {Ormes}, {Ozaki}, {Paneque}, {Panetta},
  {Parent}, {Pesce-Rollins}, {Pierbattista}, {Piron}, {Pivato}, {Porter},
  {Rain{\`o}}, {Rando}, {Razzano}, {Reimer}, {Reimer}, {Romoli}, {Roth},
  {Sada}, {Sadrozinski}, {Sanchez}, {Sbarra}, {Sgr{\`o}}, {Siskind}, {Spandre},
  {Spinelli}, {Strong}, {Suson}, {Takahashi}, {Takahashi}, {Tanaka}, {Thayer},
  {Thayer}, {Thompson}, {Tibaldo}, {Tibolla}, {Tinivella}, {Torres}, {Tosti},
  {Tramacere}, {Troja}, {Uchiyama}, {Uehara}, {Usher}, {Vandenbroucke},
  {Vasileiou}, {Vianello}, {Vitale}, {Waite}, {Wang}, {Winer}, {Wood},
  {Yamamoto}, {Yang}, \& {Zimmer}}]{fermimc}
{Ackermann}, M., {Ajello}, M., {Allafort}, A., {et~al.} 2012b, \apj, 755, 22

\bibitem[{Ackermann {et~al.}(2012)}]{pass7}
Ackermann, M. {et~al.} 2012, Astrophys.J.Suppl., 203, 4

\bibitem[{Adriani {et~al.}(2011{\natexlab{a}})Adriani, Barbarino, Bazilevskaya,
  Bellotti, Boezio, Bogomolov, Bongi, Bonvicini, Borisov, Bottai, Bruno,
  Cafagna, Campana, Carbone, Carlson, Casolino, Castellini, Consiglio,
  De~Pascale, De~Santis, De~Simone, Di~Felice, Galper, Gillard, Grishantseva,
  Jerse, Karelin, Koldashov, Krutkov, Kvashnin, Leonov, Malakhov, Malvezzi,
  Marcelli, Mayorov, Menn, Mikhailov, Mocchiutti, Monaco, Mori, Nikonov,
  Osteria, Palma, Papini, Pearce, Picozza, Pizzolotto, Ricci, Ricciarini,
  Rossetto, Sarkar, Simon, Sparvoli, Spillantini, Stochaj, Stockton, Stozhkov,
  Vacchi, Vannuccini, Vasilyev, Voronov, Wu, Yurkin, Zampa, Zampa, \&
  Zverev}]{pamelae}
Adriani, O., Barbarino, G.~C., Bazilevskaya, G.~A., {et~al.}
  2011{\natexlab{a}}, Phys. Rev. Lett., 106, 201101

\bibitem[{{Adriani} {et~al.}(2013){Adriani}, {Barbarino}, {Bazilevskaya},
  {Bellotti}, {Boezio}, {Bogomolov}, {Bongi}, {Bonvicini}, {Borisov}, {Bottai},
  {Bruno}, {Cafagna}, {Campana}, {Carbone}, {Carlson}, {Casolino},
  {Castellini}, {De Pascale}, {De Santis}, {De Simone}, {Di Felice}, {Formato},
  {Galper}, {Grishantseva}, {Karelin}, {Koldashov}, {Koldobskiy}, {Krutkov},
  {Kvashnin}, {Leonov}, {Malakhov}, {Marcelli}, {Mayorov}, {Menn}, {Mikhailov},
  {Mocchiutti}, {Monaco}, {Mori}, {Nikonov}, {Osteria}, {Palma}, {Papini},
  {Pearce}, {Picozza}, {Pizzolotto}, {Ricci}, {Ricciarini}, {Rossetto},
  {Sarkar}, {Simon}, {Sparvoli}, {Spillantini}, {Stozhkov}, {Vacchi},
  {Vannuccini}, {Vasilyev}, {Voronov}, {Yurkin}, {Wu}, {Zampa}, {Zampa},
  {Zverev}, {Potgieter}, \& {Vos}}]{pamela2}
{Adriani}, O., {Barbarino}, G.~C., {Bazilevskaya}, G.~A., {et~al.} 2013, \apj,
  765, 91

\bibitem[{Adriani {et~al.}(2011{\natexlab{b}})}]{pamela}
Adriani, O. {et~al.} 2011{\natexlab{b}}, Science, 332, 69

\bibitem[{{Aharonian}(2001)}]{FA2001}
{Aharonian}, F.~A. 2001, Space Science Reviews, 99, 187

\bibitem[{{Allen}(1973)}]{allen1973}
{Allen}, C.~W. 1973, {Astrophysical quantities}

\bibitem[{Atwood {et~al.}(2009)}]{fermi}
Atwood, W. {et~al.} 2009, Astrophys.J., 697, 1071

\bibitem[{{Bolatto} {et~al.}(2013){Bolatto}, {Wolfire}, \& {Leroy}}]{bolatto13}
{Bolatto}, A.~D., {Wolfire}, M., \& {Leroy}, A.~K. 2013, \araa, 51, 207

\bibitem[{{Casanova} {et~al.}(2010){Casanova}, {Aharonian}, {Fukui}, {Gabici},
  {Jones}, {Kawamura}, {Onishi}, {Rowell}, {Sano}, {Torii}, \&
  {Yamamoto}}]{casanova10}
{Casanova}, S., {Aharonian}, F.~A., {Fukui}, Y., {et~al.} 2010, \pasj, 62, 769

\bibitem[{{Dame} {et~al.}(2001){Dame}, {Hartmann}, \& {Thaddeus}}]{dame01}
{Dame}, T.~M., {Hartmann}, D., \& {Thaddeus}, P. 2001, \apj, 547, 792

\bibitem[{{Dame} {et~al.}(1987){Dame}, {Ungerechts}, {Cohen}, {de Geus},
  {Grenier}, {May}, {Murphy}, {Nyman}, \& {Thaddeus}}]{dame87}
{Dame}, T.~M., {Ungerechts}, H., {Cohen}, R.~S., {et~al.} 1987, \apj, 322, 706

\bibitem[{{Gabici} {et~al.}(2007){Gabici}, {Aharonian}, \& {Blasi}}]{gabici07}
{Gabici}, S., {Aharonian}, F.~A., \& {Blasi}, P. 2007, \apss, 309, 365

\bibitem[{{Grenier} {et~al.}(2005){Grenier}, {Casandjian}, \&
  {Terrier}}]{grenier05}
{Grenier}, I.~A., {Casandjian}, J.-M., \& {Terrier}, R. 2005, Science, 307,
  1292

\bibitem[{{Hillas}(2005)}]{hillas05}
{Hillas}, A.~M. 2005, Journal of Physics G Nuclear Physics, 31, 95

\bibitem[{{Kamae} {et~al.}(2005){Kamae}, {Abe}, \& {Koi}}]{kamae05}
{Kamae}, T., {Abe}, T., \& {Koi}, T. 2005, \apj, 620, 244

\bibitem[{Kelner {et~al.}(2006)Kelner, Aharonian, \& Bugayov}]{kelner}
Kelner, S., Aharonian, F.~A., \& Bugayov, V. 2006, Phys.Rev., D74, 034018

\bibitem[{{Lampton} {et~al.}(1976){Lampton}, {Margon}, \& {Bowyer}}]{fit}
{Lampton}, M., {Margon}, B., \& {Bowyer}, S. 1976, \apj, 208, 177

\bibitem[{{Montmerle}(1979)}]{montmerle79}
{Montmerle}, T. 1979, \apj, 231, 95

\bibitem[{Mori(2009)}]{mori09}
Mori, M. 2009, Astroparticle Physics, 31, 341

\bibitem[{Neronov {et~al.}(2012)Neronov, Semikoz, \& Taylor}]{nero12}
Neronov, A., Semikoz, D., \& Taylor, A. 2012, Phys.Rev.Lett., 108, 051105

\bibitem[{{Pedaletti} {et~al.}(2013){Pedaletti}, {Torres}, {Gabici}, {de
  O{\~n}a Wilhelmi}, {Mazin}, \& {Stamatescu}}]{pedaletti}
{Pedaletti}, G., {Torres}, D.~F., {Gabici}, S., {et~al.} 2013, \aap, 550, A123

\bibitem[{{Perrot} \& {Grenier}(2003)}]{gould}
{Perrot}, C.~A. \& {Grenier}, I.~A. 2003, \aap, 404, 519

\bibitem[{{Planck Collaboration} {et~al.}(2011){Planck Collaboration}, {Ade},
  {Aghanim}, {Arnaud}, {Ashdown}, {Aumont}, {Baccigalupi}, {Balbi}, {Banday},
  {Barreiro}, \& et~al.}]{planck}
{Planck Collaboration}, {Ade}, P.~A.~R., {Aghanim}, N., {et~al.} 2011, \aap,
  536, A19

\bibitem[{{Strong} {et~al.}(2007){Strong}, {Moskalenko}, \&
  {Ptuskin}}]{strong07}
{Strong}, A.~W., {Moskalenko}, I.~V., \& {Ptuskin}, V.~S. 2007, Annual Review
  of Nuclear and Particle Science, 57, 285

\bibitem[{{Strong} {et~al.}(2011){Strong}, {Orlando}, \& {Jaffe}}]{strong11}
{Strong}, A.~W., {Orlando}, E., \& {Jaffe}, T.~R. 2011, \aap, 534, A54

\bibitem[{{Villante} \& {Vissani}(2009)}]{vilante09}
{Villante}, F.~L. \& {Vissani}, F. 2009, Nuclear Physics B Proceedings
  Supplements, 188, 261

\bibitem[{{Vladimirov} {et~al.}(2011){Vladimirov}, {Digel}, {J{\'o}hannesson},
  {Michelson}, {Moskalenko}, {Nolan}, {Orlando}, {Porter}, \&
  {Strong}}]{galprop}
{Vladimirov}, A.~E., {Digel}, S.~W., {J{\'o}hannesson}, G., {et~al.} 2011,
  Computer Physics Communications, 182, 1156

\end{thebibliography}
\end{document}